\documentclass[preprint,5p,times,twocolumn,authoryear]{elsarticle}

\usepackage{hyperref}
\usepackage{mathptmx}
\usepackage{txfonts}
\usepackage{amsmath}
\usepackage[T1]{fontenc}
\usepackage{graphicx}	
\usepackage{amssymb}	
\usepackage{newtxtext,newtxmath}
\usepackage{inputenc}

\journal{Astroparticle Physics}

\begin{document}

\begin{frontmatter}



\author[1]{Ekrem O\u{g}uzhan Ang\"uner}
\ead{oguzhan.anguner@tubitak.gov.tr}
\author[2,3]{T\"ul\"un Ergin}
\affiliation[1]{country={TÜBİTAK Research Institute for Fundamental Sciences, 41470 Gebze, Turkey{~}}}
\affiliation[2]{Department of Physics and Astronomy, Michigan State University, East Lansing, MI, USA}
\affiliation[3]{Middle East Technical University, Northern Cyprus Campus, 99738 Kalkanli via Mersin 10, Turkey}
\title{The expected potential of hadronic PeVatron searches with spectral $\gamma$-ray data from the Southern Wide-field Gamma-ray Observatory}

\begin{abstract}

The presence of a spectral softening, occurring at $\sim$3~PeV energies, seen in the local cosmic-ray energy spectrum provides an evidence that our Galaxy hosts astrophysical objects, known as 'hadronic PeVatrons', that are capable of accelerating hadrons to PeV energies and above. Recent results from ground-based particle detector array experiments have provided conclusive evidence that these facilities are essential to explore the ultra-high-energy (UHE,~E$>$100 TeV) $\gamma$-ray domain and pinpoint the location of PeVatrons in the Galaxy. The Southern Wide-field Gamma-ray Observatory (SWGO) is proposed next-generation ground-based extensive air shower observatory planned for construction in the Southern Hemisphere, which holds great scientific potential for UHE observations. In this study, we investigate the expected potential of SWGO to search for hadronic PeVatrons, based on the publicly available preliminary SWGO straw-man instrument response functions (IRFs). By using these straw-man IRFs, it can be shown that the SWGO detection of $\gamma$-ray spectral cutoffs between 30~TeV and 100~TeV, at a 95$\%$ confidence level, is possible for faint $\gamma$-ray sources of $\sim$5~mCrab given that the spectral index is hard ($\Gamma\lesssim$ 2.0), while spectral cutoffs from softer sources with $\Gamma\cong$2.3 can be detected for sources brighter than $\sim$11$-$12~mCrab. The reconstructed SWGO PeVatron detection maps demonstrate that the future SWGO experiment can probe large parts of the investigated PeVatron parameter space, providing a robust detection and/or rejection of presence of spectral signatures associated with hadronic PeVatrons. A dedicated study on the promising Southern-sky PeVatron candidates, the Galactic Center region, Westerlund~1, HESS~J1702$-$420 and HESS~J1641$-$463, shows that the SWGO will have a great potential to confirm or exclude PeVatron nature of these candidate sources at a robust significance level after 5-years of observation. In addition, it is shown that controlling systematic errors will be necessary to reach full potential of the SWGO experiment for PeVatron searches.

\end{abstract}




\end{frontmatter}

\section{Introduction}
\label{intro}
Cosmic Rays (CRs) are charged particles, roaming in our Galaxy with relativistic speeds and arriving the Earth's atmosphere isotropically from outer space. Decades of measurements have shown that the composition of CRs consist mainly of protons ($\sim$90$\%$) followed by Helium nuclei ($\sim$9$\%$), while the rest are heavier ions and electrons (see \cite{blasi_rev, elena_rev2, elena_rev} for detailed review). The energy spectrum of CRs measured on Earth ranges from a few MeVs to beyond 10$^{20}$~eV. Above $\sim 30$~GeV energies, it was historically described by a smooth power-law spectrum with an index of $-$2.7 up to the so called "knee" feature, emerging at $\sim$3~PeV (1~PeV=$10^{15}$~eV), where the spectrum steepens significantly to $-$3.1 above these energies. When only the cosmic Hydrogen and Helium spectra are considered, there is an evidence that the respective knee is below 1~PeV, at energies around 700~TeV \citep{argo1PeV}. The detailed CR measurements have provided evidence of spectral hardening around 100$-$300~GeV energies and re-softening at higher energies around $\sim$100~TeV for all type of nuclei \citep{300GeV_pam,300GeV_pam2,300GeV_pam3,Adriani2022,CardilloGuliani2023}, especially for protons and Helium. These unique spectral characteristics suggest that a rigidity-dependent acceleration mechanism is at work in Galactic CR sources, allowing heavier nuclei to be accelerated to higher energies \citep{elena_rev, elena_sabrina}.~Consequently, the knee feature observed in the CR spectrum can be interpreted as a sum of different nuclei types, each showing their unique cutoff in their respective spectra. As a result, the overall CR spectrum exhibiting more or less a smooth power-law up to 3~PeV energies provides a strong evidence that the sources in our Galaxy must accelerate CRs at least up to PeV energies and even well beyond. The existence of a "second knee" feature seen at $\sim$100~PeV energies, which is thought to be originated from the heaviest nuclei, such as Fe (Z=26), further supports this idea \citep{Schroder2019,Coleman2019,pierre_rev}.~Given this scenario, it is believed that the second knee structure marks the lower bound of the energy region where transition from Galactic to the extra-galactic CRs occurs. Additionally, the recent detection of ultra-high-energy (UHE,~E$>$100 TeV) diffuse $\gamma$-ray emission from the Galactic plane \citep{tibet_diffuse,lhaaso_diffuse} offers an alternative indirect way to study the distribution and propagation of global CRs in the Galaxy, when compared to local CR observations performed on Earth. More detailed reviews on the topic of CRs can be found in \cite{elena_rev2, elena_rev, elena_sabrina, pierre_rev, Sciascio_rev, vink_rev, sabrina_rev}.

In this context, the term "PeVatron" stands for astrophysical sources which can energize CRs at least up to PeV energies and beyond. Various astrophysical source classes, such as supernova (SN) Remnants (SNRs) \citep{bell1978,GabiciAharonian07, celli_SNR}, massive star clusters (MSCs) \citep{aharonian2019, morlino_sc}, core-collapse SNe \citep{tatischeff2009,bell2013,zirakashvili2016,Marcowith2018}, pulsar wind nebulae (PWNe) \citep{pulsarsElena1,Ohira2018,pulsarsClaire3,pulsarsElena2}, TeV halos \citep{lindenTeVHalo, fangTeVHalo}, star formation regions (SFRs) \citep{sfr_bykov}, microquasars \citep{microquasars} and superbubbles \citep{higdon2003,binns2005,Vieu2022}, are expected to be promising PeVatron candidates. PeVatrons are categorized as "hadronic PeVatrons" or "leptonic PeVatrons" depending on whether the accelerated particles are primarily hadrons or leptons, respectively. The definition of PeVatron focuses on the maximum achievable energy of an accelerator, identifying the PeVatron nature based on the highest energy reached by individual particles at the acceleration site. This definition establishes a strict energy threshold of '1~PeV', distinguishing between PeVatron and non-PeVatron nature of the source, consequently making the 1~PeV energy threshold the primary property of a PeVatron. The detection of UHE $\gamma$-rays above 1~PeV from the Crab Nebula \citep{crab_pev}, a known host of leptons with PeV energies for over a decade \citep{AmatoOlmi21}, and the Cygnus Cocoon region \citep{lhaaso2021}, a superbubble surrounding a massive star formation region \citep{cygnus_cocoon}, implies the existence of particles accelerated to PeV energies at these sites, regardless of whether the emission has hadronic or leptonic origin, making the Crab Nebula and Cygnus Cocoon robust Galactic PeVatron sources according to the definition given above. The question of the origin of CRs, specifically the astrophysical accelerators capable of energizing CRs up to the observed knee feature around $\sim$3~PeV energies, stands as one of the top scientific questions in astrophysics. Within this context, the investigation into the origin of CRs is directly linked to hadronic PeVatrons, as the contribution of leptons to the observed CR spectrum is negligible ($\sim$1-2$\%$). To account for the knee feature around $\sim$3~PeV, CR sources must efficiently accelerate bulk of hadrons well beyond 1~PeV energies. Consequently, the hadronic spectrum produced by the accelerator should extend beyond 1~PeV energies without exhibiting a spectral cutoff.

From the experimental point of view, observed high-energy (HE,~E$>$100~MeV) and UHE $\gamma$-ray emission are the primary means to explore the origins of CRs. When $\gamma$-rays are created from proton-proton (pp) interactions followed by subsequent pion decay, the energy of $\gamma$-rays is typically around 10 times lower than the energy of parent protons \citep{kernel2006, kafexhiu2014, celli_2020}, and can create spectral features in the $\gamma$-ray spectrum, providing clear signatures of hadronic interactions. For example, the AGILE and Fermi-LAT results \citep{Giuliani2011,Ackermann2013} demonstrated that Galactic SNRs can accelerate CRs, producing low energy (MeV$-$GeV) $\gamma$-rays showing up as a "pion-bump" feature\footnote{The pion bump is a characteristic spectral feature observed in the spectral energy distribution of $\gamma$ rays, seen in between 100~MeV and 1~GeV energies. This feature is a result of hadronic interactions between accelerated CRs with the surrounding gas. Please refer to \cite{pion_bump} for more details.} in the observed SNRs' energy spectra. On the other hand, regardless of whether they are hadronic or leptonic, significant UHE emission is one of the main characteristics of PeVatron sources and must be detected for a robust claim of PeVatron nature. In the hadronic case, correlation with target material, in which accelerated particles can interact with, is expected. It is important to note that the target region, where CR interactions take place, is an evidence of PeVatron activity, but the region does not necessarily host the PeVatron source itself. In addition, neutrino emission is a clear indication of hadronic interactions \citep{kernel2006, neutrino1}. However, the sensitivities of current neutrino experiments,~i.e.~IceCube \citep{IceCube_Abbasi2014} and Baikal-GVD \citep{BaikalGVD_Abbasi2014}, are not sufficient to put strong constraints on significant discrimination between hadronic and leptonic nature of PeVatrons \citep{IceCubeGalMap}, while the future neutrino observatories, such as IceCube-Gen2\footnote{https://www.icecube-gen2.de/index$_{-}$eng.html} and KM3NeT\footnote{https://www.km3net.org/}, are expected to reach such high sensitivities.

The first indication of the presence of a Galactic PeVatron at the Galactic center (GC) region is discussed in \cite{hess_gc_pevatron}. Detection of the GC PeVatron was claimed based on the derived 95$\%$ confidence level (CL) lower limits on the parent proton spectral cutoff of 0.4~PeV, together with the strong correlation observed between the very-high energy (VHE, 0.1~TeV$<$E$<$100~TeV) $\gamma$-ray emission and distribution of molecular gas. After this pioneering study, interpreting PeVatron nature of a source based on 95$\%$ CL lower limits of the proton spectral cutoff became a standard approach in PeVatron searches \citep{gerrit_pevatrons,j1702}. In the following years, ground-based water Cherenkov detector (WCD) experiments, Tibet AS-gamma \citep{tibet_first,tibet_snr} and the High-Altitude Water Cherenkov Observatory (HAWC) \citep{hawc_E56TeV}, reported significant detection of Galactic UHE photons and sources, respectively. Finally, the discovery of 12 Galactic UHE source by the LHAASO collaboration marked a major milestone in PeVatron searches, and completely reshaped our understanding of the PeVatron concept, as many of these sources are plausibly associated to leptonic accelerators like pulsars and PWNe \citep{lhaaso2021}. The recently published first LHAASO catalogue of $\gamma$-ray sources have revealed 43 UHE sources detected above 100~TeV with significance greater than 4$\sigma$ \citep{lhaaso_1stcat}. 

These experimental results have provided clear evidence that ground-based WCDs, e.g.~HAWC and LHAASO, due to their enhanced high energy flux sensitivities above $\sim$10~TeV, are the key facilities to explore the UHE regime and locate PeVatrons in the Galaxy, therefore making them "PeVatron hunters". However, up to date, no Galactic source that has been firmly proven to accelerate hadrons to PeV energies and above could be identified. This is due to the fact that discrimination between hadronic and leptonic PeVatrons, using current observational data, are extremely challenging. On the other hand, angular resolution of WCD experiments ($\sim$0.2$^\circ$-0.3$^\circ$) is a factor of 3$-$4 worst when compared to current generation imaging atmospheric Cherenkov telescopes (IACTs). Such a disadvantage causes source confusion to become a major problem for the ground-based WCD observations, consequently making it impossible to pin down the astrophysical origin of the observed emission from morphology studies. The synergy between future VHE and UHE $\gamma$-ray instruments such as the Cherenkov Telescope Array (CTA) \citep{Acharya2013,HofmannZanin2023} and the Southern Wide-field Gamma-ray Observatory (SWGO) \citep{swgo_white} will play a critical role for robust identification of Galactic PeVatrons which can contribute significantly to the CR knee feature. In addition, future neutrino experiments, like KM3NeT and IceCube-Gen2, will be able to provide sensitive measurements that can resolve the ambiguity between leptonic and hadronic PeVatrons. More detailed reviews on the topic of PeVatrons can be found in \cite{elena_sabrina,elena_rev,elena_rev2,Sciascio_rev,sabrina_rev,pierre_rev, vink_rev, sudoh_pevatrons, Fuente2023ThePA, CardilloGuliani2023, ozi_review, pev_annual_rev}.

Throughout this paper, we focus on the hadronic PeVatrons, namely source of CRs at energies around the knee of the CR spectrum. Consequently, the leptonic PeVatrons are out of the scope of this study. The paper is structured as follows. The next-generation SWGO experiment is briefly introduced in Sec.~\ref{swgo_sec}. Simulations of the SWGO observations and the data analysis methods employed in this study are discussed in Sec.~\ref{sim_analyze}. The results on the investigation of SWGO’s sensitivity in $\gamma$-ray spectral cutoff detection are provided and discussed in Sec.~\ref{swgo_cutoff_sens}. The general ability of SWGO to identify PeVatron sources is quantified in Sec.~\ref{sec_par_scan}. Simulation results of the SWGO observations of the promising PeVatron candidates in the Southern-sky are provided and discussed in Sec.~\ref{promising_sources}. Finally, the discussions and conclusions are summarized in Sec.~\ref{discussions}.

\section{The Southern Wide-field Gamma-ray Observatory}
\label{swgo_sec}
SWGO is a proposed next-generation ground-based Extensive Air Shower (EAS) observatory to be built in the Southern Hemisphere, designed to scan large parts of the sky with a large field of view (FoV) at a very high duty cycle ($>$95$\%$), and measure $\gamma$-rays within the range spanning from a few hundred GeVs to PeV energies \citep{swgo_white}. Expected to start full operations by 2026, SWGO is set to maintain its operational capacity for a minimum of 10 years. The fundamental structure of SWGO consists of ground-based WCDs, positioned in the Southern Hemisphere at an altitude exceeding 4400 m above sea level \citep{SwgoConceicao2023,SwgoBarresdeAlmeida2021}. These WCDs are designed to optimize and enhance both signal collection and particle recognition capabilities through the exceptionally precise time resolution ($\Delta$t $\approx$ 2 ns) and advanced signal reconstruction techniques, respectively. One of the proposed detector designs is a dual-layered water Cherenkov tank, which will enable reconstructing the number of muons and separating primary CRs, therefore both increasing the background rejection rate and allow CR anisotropy measurements \citep{SwgoLang2023}, respectively.

The scientific potential and capabilities of ground-based EAS arrays have already been demonstrated by the current generation observatories like HAWC \citep{hawc_sens} and LHAASO \citep{lhaaso_sens}, both of which are located in the Northern Hemisphere. Despite the Southern-sky holds great scientific potential, there is currently no operational EAS observatory in the Southern Hemisphere. In this regard, SWGO will serve as a complementary to the currently existing ground-based EAS observatories in the Northern Hemisphere, and to the next-generation IACT project, the CTA observatory. Through a five-year data collection with SWGO, it is expected that the sensitivity of detecting point-like $\gamma$-ray sources at energies exceeding 10 TeV will be better when compared to 50 hours of CTA data \citep{swgo_white}. This advantage, together with its large FoV, makes SWGO an ideal observatory to search for $\gamma$-ray sources above several 10 TeV. Such capability is particularly important for revealing sources of Galactic CR acceleration and the systematic investigation of transient events at these high energies. Furthermore, SWGO has the potential to play a key role in subsequent follow-up observation campaigns in case of transient alerts and multi-messenger triggers, both by mapping the distribution of transient and by extending the simultaneous sky coverage of $\gamma$-ray monitoring facilities, e.g.~HAWC, LHAASO, and CTA \citep{LaMura2020}, respectively. In addition, strategic location of SWGO close to the Southern Tropic \citep{swgo_white} ensures efficient coverage of the declination band of transient events that might be associated to neutrino alerts from current neutrino observatories, e.g. IceCube\footnote{https://icecube.wisc.edu/science/real-time-alerts/} \citep{IceCube_alerts}, future neutrino observatories like KM3NeT and IceCube-Gen2, and gravitational wave observatories, such as LIGO \footnote{https://emfollow.docs.ligo.org/userguide/early$_{-}$warning.html} \citep{LIGO_alerts}.

The search for the location of the SWGO construction site has been in progress as a part of the research and development phase, and is expected to reach its conclusion by the end of 2024. Up to date, the SWGO collaboration has been systematically gathering detailed information regarding the potential sites under consideration, which are located in Argentina, Chile, and Peru \citep{SwgoDoro2021}. The selection of these candidate sites was based on various factors including altitude, local topology, environmental conditions, site access, transport costs, as well as the availability and cost of essential resources like water, power, and network connectivity \citep{SwgoSantander2023}. 

\begin{figure}[!ht]
\includegraphics[width=9.0cm]{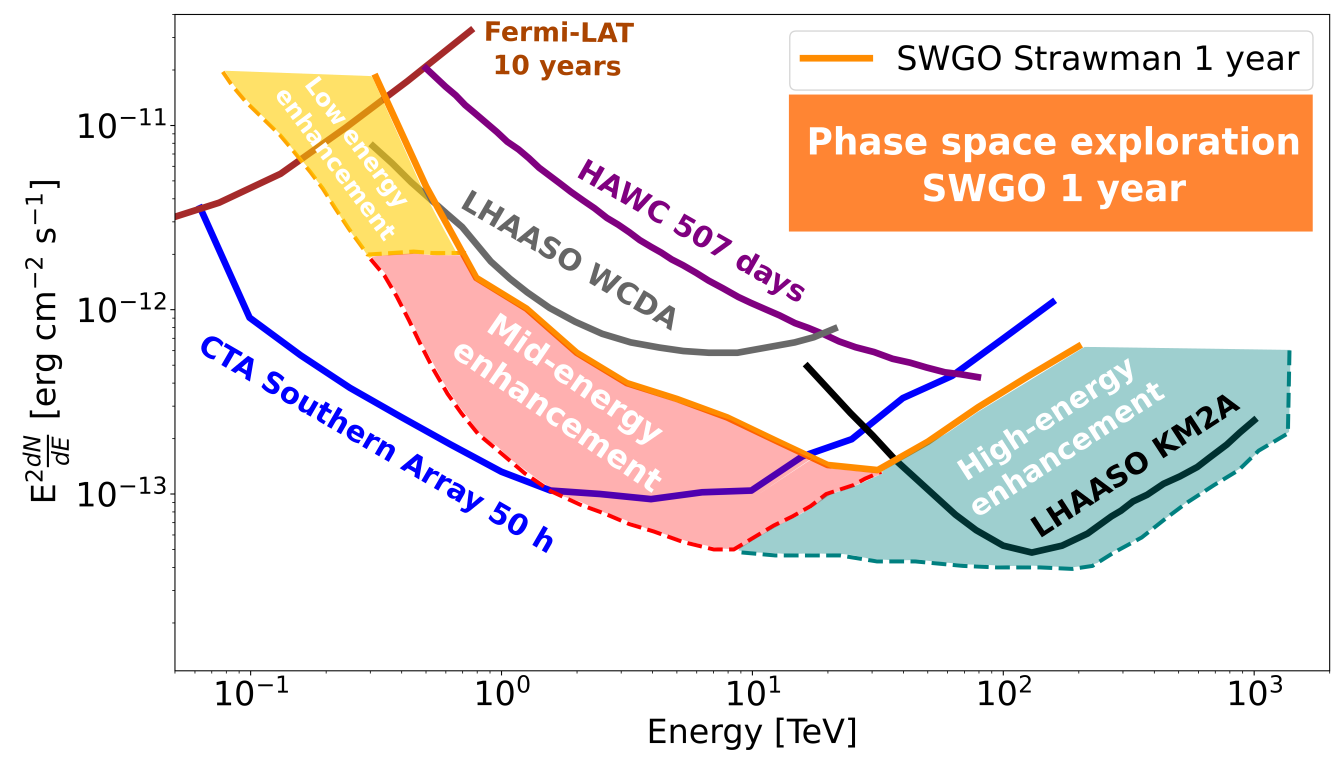}
\caption{A comparison of differential point-like source sensitivities between the SWGO (the 1 year straw-man, solid-orange, taken from \cite{swgo_white}), Fermi-LAT (10 years P8R3$_{-}$SOURCE$_{-}$V3, solid-brown, taken from \cite{fermi_10years}), CTA (50 h Southern array prod5.v0.1, solid-blue, taken from \cite{prod5}), HAWC (507 days, solid-magenta, taken from \cite{hawc_sens}) and LHAASO~WCDA (1 year, solid-gray) and LHAASO~KM2A (1 year, solid-black) taken from \cite{lhaaso_sensitivity, Volpe_2023}. The yellow, red and blue shaded bands indicate the foreseen phase-space exploration of the low-energy, mid-energy and high-energy enhancement, respectively, for the SWGO experiment. The figure is adapted from \cite{jim_swgo}.}
\label{fig_swgo_sens}
\end{figure}

Figure~\ref{fig_swgo_sens} shows the differential point-like source sensitivity curves from various experiments, along with the sensitivity of the SWGO for 14 different detector and array layout configurations, which are indicated with shaded bands featuring different colors. The selection of these configurations follows a detailed investigation of array and detector parameters, including dimensions of the detector station, number and size of photo-sensors placed inside the detector, and the distribution of the dense inner (for lower energies) and sparse outer array (for higher energies), as well as the correspondence between them. The solid orange sensitivity curve shown in Fig.~\ref{fig_swgo_sens} represents the preliminary baseline configuration of SWGO, which is referred as the \textbf{straw-man design} throughout this paper. This straw-man design configuration comprises a compact inner array with a 160 m radius and a fill factor of $\sim$80$\%$, which is surrounded by a less dense outer array, having a radius of 300 m and a fill factor of $\sim$5$\%$. The detector units in both the inner and outer arrays, resembling a two-compartment cylindrical tank with a diameter of 3.8 m, are identical and resulting in a total of 6600 detector units \citep{SwgoSchoorlemmer2021}. The sensitivity of the straw man design was evaluated between 20~GeV up to 500~TeV by extrapolating the published HAWC performance metrics \citep{hawc_sens}, such as the angular and energy resolution to be $\sim$0.15$^{\circ}$ and $\sim$25$\%$ at 30 TeV, respectively, as well as the passing rate of the gamma and hadron cut to be $\sim$2$\times10^{-3}$ (see Fig.~11 of \cite{hawc_sens}).

The sensitivity curve of SWGO shown in Fig.~\ref{fig_swgo_sens} serves as a guiding reference to direct the design studies, rather than being a definite measure of SWGO performance. The comprehensive exploration of the complete parameter phase-space is expected to be completed in 2024. The differently colored shaded regions in Fig.~\ref{fig_swgo_sens} represent various design options aimed at improving the sensitivity of the straw man design. Potential enhancements in low-energy performance (shaded yellow region) are achievable by lowering the individual unit thresholds and the exploration of higher elevation sites. In the mid-energy range (shaded red region), it has been shown that significant enhancements in both angular resolution and background rejection can be achieved \citep{hofmann2020,Kunwar2021,ruben2021}. The lower limit of the color band, marked by the dashed line, corresponds to a 30$\%$ improvement in the point spread function (PSF) and a tenfold increase in background rejection efficiency. Particularly, ongoing investigations focus on compact detector units with dedicated muon tagging capabilities to further enhance background rejection efficiency. For the high-energy range (shaded blue region), performance improvements can be achieved by implementing a low-density, large outer array with a size of a few square kilometers, coupled with effective background rejection capabilities. As it was evident from the LHAASO results \citep{lhaaso2021}, it is possible to implement a square kilometer array with a background efficiency of $\sim$10$^{-5}$. These studies will provide valuable insights for identifying the most favorable SWGO design configurations to consider, subsequently followed by the production of the official SWGO instrument response functions (IRFs). ~Therefore, we explicitly mention that the results and conclusions presented throughout this paper are not based on any official IRFs or tools provided by the SWGO collaboration. Additionally, the results and conclusions derived in this paper are conservative with regard to their dependence on the assumed differential sensitivity.

\section{Simulations and data analysis}
\label{sim_analyze}
The expected potential of identifying point-like PeVatron sources using the forthcoming SWGO experiment data is evaluated through Monte-Carlo (MC) simulations, based on the differential straw-man sensitivity\footnote{The differential straw-man sensitivity curve data are taken from https://github.com/harmscho/SGSOSensitivity.} outlined in \cite{swgo_white}. This straw-man SWGO sensitivity curve is provided with an energy binning approach. Each bin, indexed with $i$ and centered at the energy value of $E_i$, in principle, determines the minimum detectable $\gamma$-ray flux level $\Phi_\mathrm{Sens}(E_i)$ within it, ensuring a $5\sigma$ detection significance, assuming observation times of both 1 year and 5 years. For each specific $E_i$, the simulated $\gamma$-ray flux points $\Phi(E_i)$ are generated based on SWGO straw-man sensitivity curve with $\sigma(E_i)=\Phi_\mathrm{Sens}(E_i)/5$ as the standard deviation, and are distributed normally around $\Phi_\mathrm{True}(E_i)$ representing the predicted $\gamma$-ray spectrum from astrophysical sources. Spectral data points exhibiting a relative error $\sigma(E_i)/\Phi(E_i)$ exceeding 100$\%$ are excluded from subsequent analyses. This exclusion is due to the intention of deriving flux upper limits for such data points in practical analysis applications.

In the subsequent sections, either the simulated SWGO $\gamma$-ray flux data or publicly available spectral $\gamma$-ray flux data obtained from H.E.S.S. observations of various Galactic PeVatron candidate sources are analyzed within the framework of {\tt gammapy} \citep{gammapy_v018}. The analysis procedure is based on fitting the respective flux data sets to $\gamma$-ray emission models. The model parameters that best describe the data are determined by minimizing the $\chi^2$ statistic. 

When dealing with the publicly available H.E.S.S. flux data, which may include asymmetric statistical errors represented by $[\sigma_-,\sigma_+]$, the symmetric statistical errors given by $\sigma_\mathrm{stat}=\mathrm{max}\{\sigma_-,\,\sigma_+\}$ are conservatively adapted instead of directly incorporating the asymmetric errors. Additionally, relative systematic flux error of $\sigma_\mathrm{sys}=\xi\,dN/dE$ with $\xi$=20$\%$ is taken into account for the H.E.S.S. data \citep{hess_crab}. To explore the potential impact of systematic errors on the PeVatron searches with SWGO, relative systematics of $\xi$=5$\%$ (optimistic case) and $\xi$=10$\%$ (conservative case) are assumed in the simulated SWGO flux points. These assumptions are based on the 7$\%$ flux systematics observed in the LHAASO experiment \citep{lhaaso_sens_crab}. Finally, the overall error associated with the analyzed flux data points is determined by selecting the larger value as $\sigma=\mathrm{max}\{\sigma_\mathrm{sys},\,\sigma_\mathrm{stat}\}$. This combined error estimation ensures a conservative consideration of the uncertainties in the analysis.

The differential spectrum of $\gamma$-ray sources are modeled as exponential cutoff power law (ECPL) model formulated as follows 
\begin{equation}
\label{eq1}
\Phi_\mathrm{ECPL}(E)=\Phi_{0}\mathrm{(E_{0})} \cdot \left(\frac{E}{E_{0}}\right)^{-\Gamma_{\gamma}}
\cdot \exp{\left(-\lambda_{\gamma}\;E \right)}\,\mathrm{,}
\end{equation}
where $\lambda_{\gamma}$=(1/$E_{cut,\,\gamma}$) is the inverse $\gamma$-ray cutoff energy with $E_{cut,\,\gamma}$ representing the cutoff energy of $\gamma$-ray spectrum, $\Gamma_{\gamma}$ is the spectral index and $\Phi_{0}$($E_{0}$) is the source flux normalization at the reference energy $E_{0}$. Likewise, the differential energy distribution of accelerated protons is also assumed to follow an ECPL model expressed as 
\begin{equation}
J_\mathrm{p}(E_\mathrm{p}) \sim E_\mathrm{p}^{-\Gamma_\mathrm{P}}\;\exp\left(-\left(\lambda_{\mathrm{p}}\;{E_\mathrm{\mathrm{p}}}\right)^{\beta}\right)\,\mathrm{,}
\label{eq2}
\end{equation}
where $\lambda_{\mathrm{p}}$=(1/$E_\mathrm{cut,\,\mathrm{p}}$) is the inverse proton cutoff energy $E_\mathrm{cut,\,p}$ and $\Gamma_\mathrm{P}$ is the proton spectral index. The parameter $\beta$ describes the degree of sharpness in the exponential cutoff, and for the analyses presented in this paper, it is fixed to $\beta$=1. This choice adequately captures the particle spectra characteristics in scenarios involving hadronic acceleration as discussed in \cite{Schure13, Cristofari20, pts_paper}. The likelihood test statistics, 
\begin{equation}
\label{ts_lambda}
\mathrm{TS}_{\gamma,\mathrm{p}}=-2\ln\frac{\hat{L}(\lambda_{\gamma,\mathrm{p}}=0)}{\hat{L}(\lambda_{\gamma,\mathrm{p}})},
\end{equation}
where $\hat{L}(\lambda_{\gamma,\mathrm{p}})$ and $\hat{L}(\lambda_{\gamma,\mathrm{p}}=0)$ are the maximum likelihoods over the full parameter space, either for $\gamma$-rays ($\lambda_{\gamma},\,\Phi_{0,\gamma},\,\Gamma_{\gamma}$) or protons $(\lambda_{\mathrm{p}},\,\Phi_{0,\mathrm{p}},\,\Gamma_{\mathrm{p}})$, used for quantifying the statistical significance of $\gamma$-ray spectral energy cutoffs ($\mathrm{S}_\mathrm{cut,\gamma}$=$\sqrt{\mathrm{TS}_\mathrm{\gamma}}$) and protons spectral cutoffs ($\mathrm{S}_\mathrm{cut,p}$=$\sqrt{\mathrm{TS}_\mathrm{p}}$), respectively. The PeVatron Test Statistics (PTS) method, as introduced in \cite{cta_pevatrons} and formulated as 
\begin{equation}
\label{eq_PTS}
\mathrm{PTS}=-2\ln\frac{\hat
L(E_\mathrm{cut,\,p}=1\,\mathrm{PeV},\boldsymbol{\theta}|D)}{\hat L(E_\mathrm{cut,\,p}, \boldsymbol{\theta}|D)}\,\mathrm{,}
\end{equation}
offers a likelihood ratio test that enables the measurement of the deviation of the best-fit hadronic energy cutoff, denoted as $E_\mathrm{cut,\,p}$, extracted from a specific set of observed data ($\mathrm{D}$), from a fixed proton cutoff energy threshold set at 1~PeV. Throughout this paper, only flux data $\Phi(E_i)$ with errors $\sigma(E_i)$ in energy bins $E_i$ are analyzed, and the adopted likelihood function is given by
\begin{equation}
L(E_\mathrm{cut,\,p},\,\boldsymbol{\theta}|D)=-2\sum_i \left(\frac{\Phi_\gamma(E_i|E_\mathrm{cut,\,p},\,\boldsymbol{\theta})-\Phi(E_i)}{\sigma(E_i)}\right)^2\,\mathrm{,}
\end{equation}
where $\boldsymbol{\theta}=(\Gamma_\mathrm{P},\Phi_{0,\mathrm{p}})$. This paper consistently employs the statistical significance of the PTS ($S_\mathrm{PTS}=\mathrm{sign}(E_\mathrm{cut,\,p}-1\,\mathrm{PeV})\sqrt{\mathrm{PTS}}$) to gauge the level of statistical significance regarding the identification of spectral PeVatron signatures. The {\tt ecpli} package~\citep{ecpli} is used to derive lower limits for both hadronic and $\gamma$-ray cutoff energies associated to a given source, following the methods as explained comprehensively in the appendix provided in \cite{cta_pevatrons}. The analysis presented in this paper does not include the impact of the attenuation of $\gamma$-ray emission due to pair creation, i.e.~the process $\gamma\gamma\rightarrow e^+ e^-$. This exclusion is due to the focus on simulated spectral $\gamma$-ray flux points with energies below 200$-$300~TeV. As outlined in \cite{absorption_vernetto}, it is assumed that the flux attenuation due to pair creation is negligible ($<$ 10$\%$) below these energies. Nevertheless, it is important to mention that when simulating SWGO data without accounting for systematic errors ($\xi=0$), the impact of pair creation can become significant, particularly depending on the location of the source in the Galaxy and its distance.

\section{Investigating SWGO's sensitivity in $\gamma$-ray spectral cutoff detection}
\label{swgo_cutoff_sens}

The SWGO Collaboration has defined a set of science benchmarks, encompassing the key target scientific objectives of the SWGO project, with the aim of exploring performance parameters that are crucial for the best possible optimization of the forthcoming SWGO experiment \citep{jim_swgo}. One of them is directly related to PeVatron searches and defined under the 'Galactic accelerators' science case. The benchmark is described as "Maximum exponential-cutoff energy detectable at 95$\%$ CL~in 5 years for a $\gamma$-ray source with spectral parameters of $\Phi_{0}$(1~TeV)=5~mCrab\footnote{The Crab unit is taken as the differential Crab flux level at 1~TeV of 3.84~$\times$ 10$^{-11}$ cm$^{-2}$ s$^{-1}$ TeV$^{-1}$ following Table 6 of \cite{hess_crab}} and $\Gamma_{\gamma}$=2.3" \citep{jim_swgo}. In this section, results of a dedicated simulation study utilizing 5-years straw-man SWGO sensitivity curve are presented. The aim of this study is to explore and provide an estimation of the maximum detectable energy cutoff at a 95$\%$ CL over a 5-year observation time as defined in the respective SWGO science benchmark. 

\begin{figure}[ht!]
\includegraphics[width=9cm]{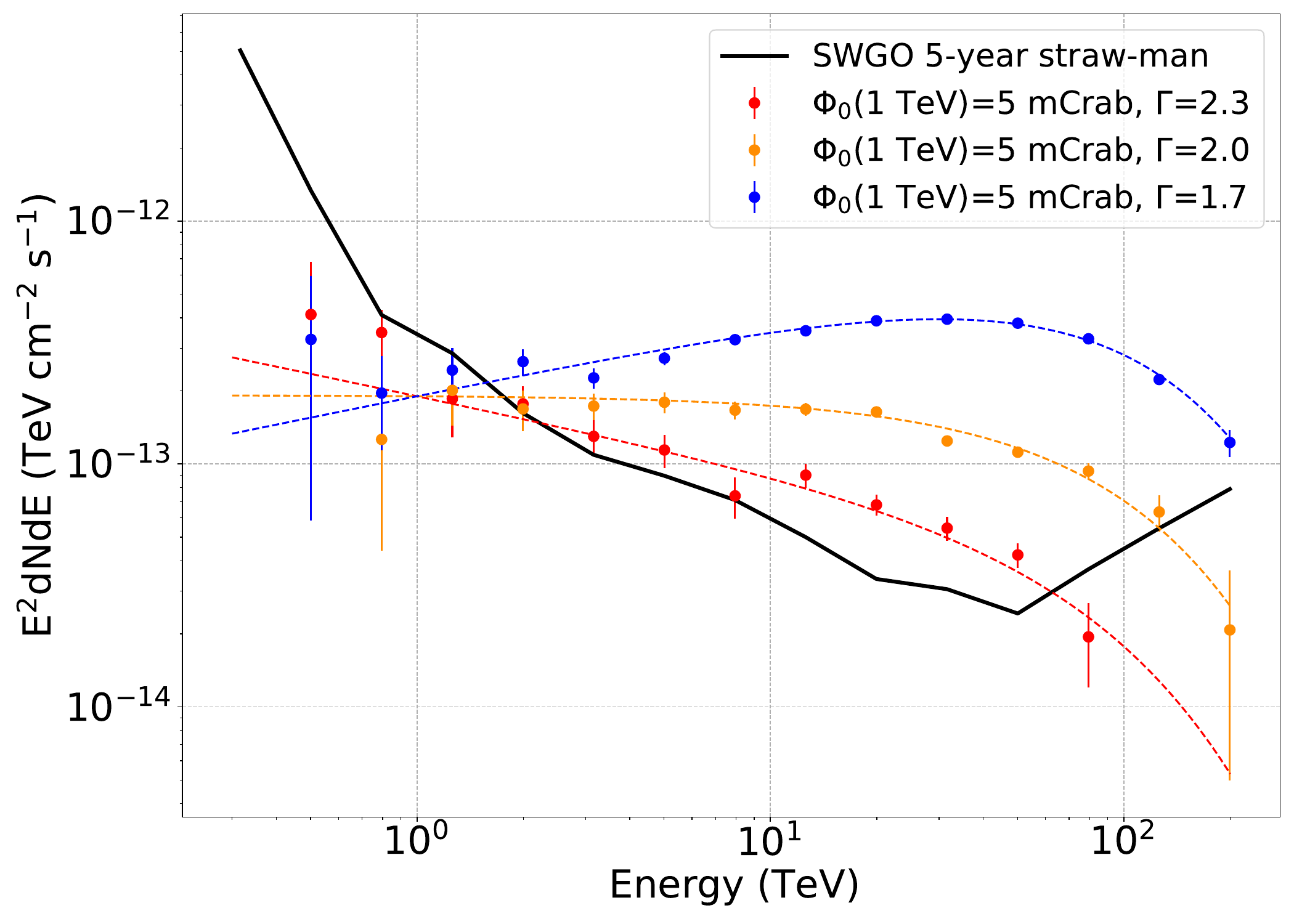}
\caption{A comparison is shown between $\gamma$-ray spectral models with different spectral indices of $\Gamma$=2.3 (red), $\Gamma$=2.0 (yellow) and $\Gamma$=1.7 (blue), for a source exhibiting a flux normalization $\Phi_{0}$ of 5~mCrab at 1~TeV, and a fixed $\gamma$-ray spectral cutoff energy of 100~TeV. Additionally, the flux points derived from SWGO simulations, following the procedure detailed in Sec.~\ref{sim_analyze}, are shown in their corresponding colors, accounting only for 1$\sigma$ statistical errors ($\xi=0$). The solid black line outlines the 5-year SWGO straw-man sensitivity curve.}
\label{fig_fluxmodels}
\end{figure}

\begin{figure*}[ht!]
\centering
\includegraphics[width=\textwidth]{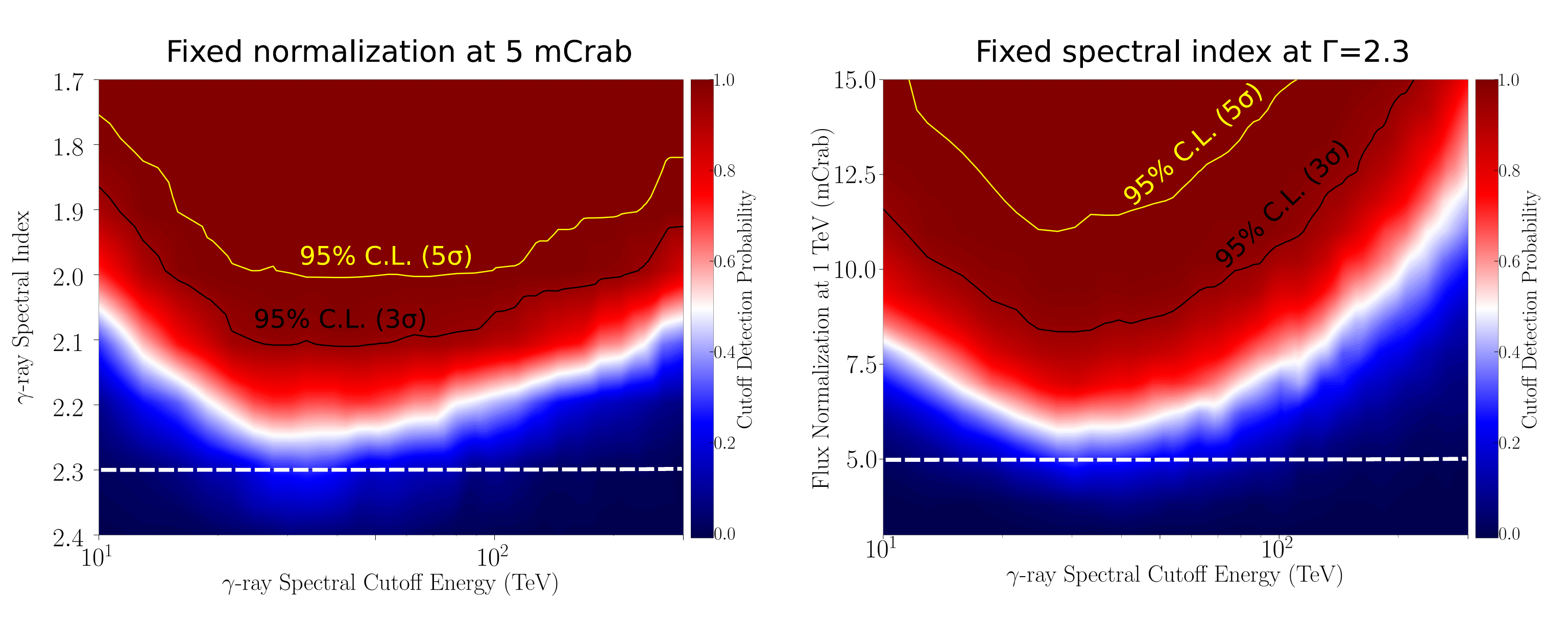}
\caption{The $\gamma$-ray spectral cutoff detection maps reconstructed from 5-years of SWGO observations are shown for a fixed flux normalization of $\Phi_{0}$(1~TeV)=5~mCrab (left) and for a fixed gamma-ray spectral index of $\Gamma$=2.3 (right). The x-axes show$\gamma$-ray spectral cutoff energies, while the y-axes show the $\gamma$-ray spectral index (left) and flux normalization at 1 TeV (right) of the simulated ECPL model given in Eq.~\ref{eq1}, respectively. For the reconstruction of the cutoff energy axis, a total of 29 equally spaced logarithmic bins between 10 and 300~TeV energies have been used. As for the spectral index and flux normalization axes, eight bins cover the range of $\Gamma$=[1.7,~2.4], and 13 bins cover the flux range of $\Phi_{0}$(1~TeV)=[3.0,~15.0]~mCrab, respectively. Within each ($\rm{\Phi_{0}}$, $\Gamma$, E$_{c,\gamma}$) combination bin, a set of 500 simulations of SWGO flux points is performed based on the corresponding $\gamma$-ray model, and the distribution of $\mathrm{TS}_{\gamma}$ is generated. The detection probabilities of the spectral cutoffs, represented on the z-axis, are calculated by assuming a cutoff detection threshold of $\mathrm{TS}_{\gamma}$ $\geq$ 25 (5 $\sigma$) and taking the fraction of the distribution above this threshold, as initially introduced in \citep{cta_pevatrons}. The yellow and black contours indicate lower bounds of the parameter space in which the 95$\%$ CL cutoff detection can be established assuming the cutoff detection threshold of $\mathrm{S}_\mathrm{cut,\gamma}$=5$\sigma$ and $\mathrm{S}_\mathrm{cut,\gamma}$=3$\sigma$, respectively. The white dashed lines in figures indicate location of the reference source cited in SWGO science benchmark, with parameters of $\Phi_{0}$(1~TeV)=5~mCrab and $\Gamma$=2.3, on the phase space.}
\label{fig_cutoffdet}
\end{figure*}

In principle, detecting a $\gamma$-ray spectral cutoff, or equivalently, robust determination of a spectral shape, requires reasonably broad energy coverage and sufficient event statistics within that energy range. Figure \ref{fig_fluxmodels} illustrates a comparison between different $\gamma$-ray spectral models with various spectral indices, originating from a 5~mCrab source exhibiting a 100~TeV $\gamma$-ray spectral cutoff, together with the simulated respective SWGO flux points following the method detailed in Sec.~\ref{sim_analyze}. The figure clearly reveals that detecting a 100~TeV $\gamma$-ray cutoff from a 5~mCrab and $\Gamma$=2.3 source (dashed red line) encounters challenges due to both insufficient statistics and the limited ability of the SWGO 5-year straw-man sensitivity curve to effectively capture the cutoff feature. Conversely, a source with a comparable flux level but showing a harder spectral index of $\Gamma$=1.7 benefits from comprehensive coverage, making the detection of a 100~TeV cutoff feature possible. A preliminary investigation of the simulated 5 mCrab and $\Gamma$=2.3 source did not conclusively result in a detection of any spectral cutoff feature up to energies of 300~TeV at a 95$\%$ CL. Consequently, the simulation study is extended further to production of spectral detection maps, introduced in \cite{cta_pevatrons}, which encompass a wide range of the $\gamma$-ray spectral parameter space and can be used to investigate the detectable cutoff energies at a desired CL.

Figure~\ref{fig_cutoffdet} shows the reconstructed $\gamma$-ray spectral cutoff detection map using 5-years of SWGO observations for a fixed flux normalization of $\Phi_{0}$(1~TeV)=5~mCrab (left) and a fixed $\gamma$-ray spectral index of $\Gamma$=2.3, together with 95$\%$ CL~contour lines assuming $\mathrm{S}_\mathrm{cut,\gamma}$=5$\sigma$ (yellow) and $\mathrm{S}_\mathrm{cut,\gamma}$=3$\sigma$ (black) cutoff detection thresholds given in Eq.~\ref{ts_lambda}, respectively. As evident from the figures, the reference source cited in the SWGO science benchmark fails to achieve a 95$\%$ CL detection for any spectral cutoff value spanning from 10 to 300~TeV when taking the straw-man design sensitivity curve into account. By analyzing the characteristics of the reference source parameters separately, it becomes apparent that $\gamma$-ray spectral cutoff energies ranging from 30 to 100~TeV can be confidently detected (at 5$\sigma$ level) for a 5~mCrab source, given that the $\gamma$-ray spectral index is hard ($\Gamma$~$\leq$~2.0). The probabilities of detecting these cutoff energies exhibit a relatively flat structure between 30~TeV and 100~TeV. On the other hand, when considering $\gamma$-ray sources with a spectral index of $\Gamma$=2.3, robust detection of their spectral cutoffs can become possible only if the flux $\Phi_{0}$(1~TeV) is larger than $\sim$11~mCrab. In this case, the capability to identify spectral cutoffs attains its peak performance at E$_{c,\gamma}$=$\sim$30~TeV, and the maximum energy at which a cutoff can be detected increases as the source gets brighter. If a less strict cutoff detection threshold of $\mathrm{S}_\mathrm{cut,\gamma}$=3$\sigma$ is assumed, the minimum prerequisites for spectral parameters become less conservative, enabling the potential detection of spectral cutoffs even for a 5~mCrab source with $\Gamma$=$\sim$2.1, as well as for a $\sim$8~mCrab and $\Gamma$=2.3 source. 

\section{Exploration of PeVatron parameter space}
\label{sec_par_scan}
\begin{figure*}[ht!]
\centering
\includegraphics[width=17cm]{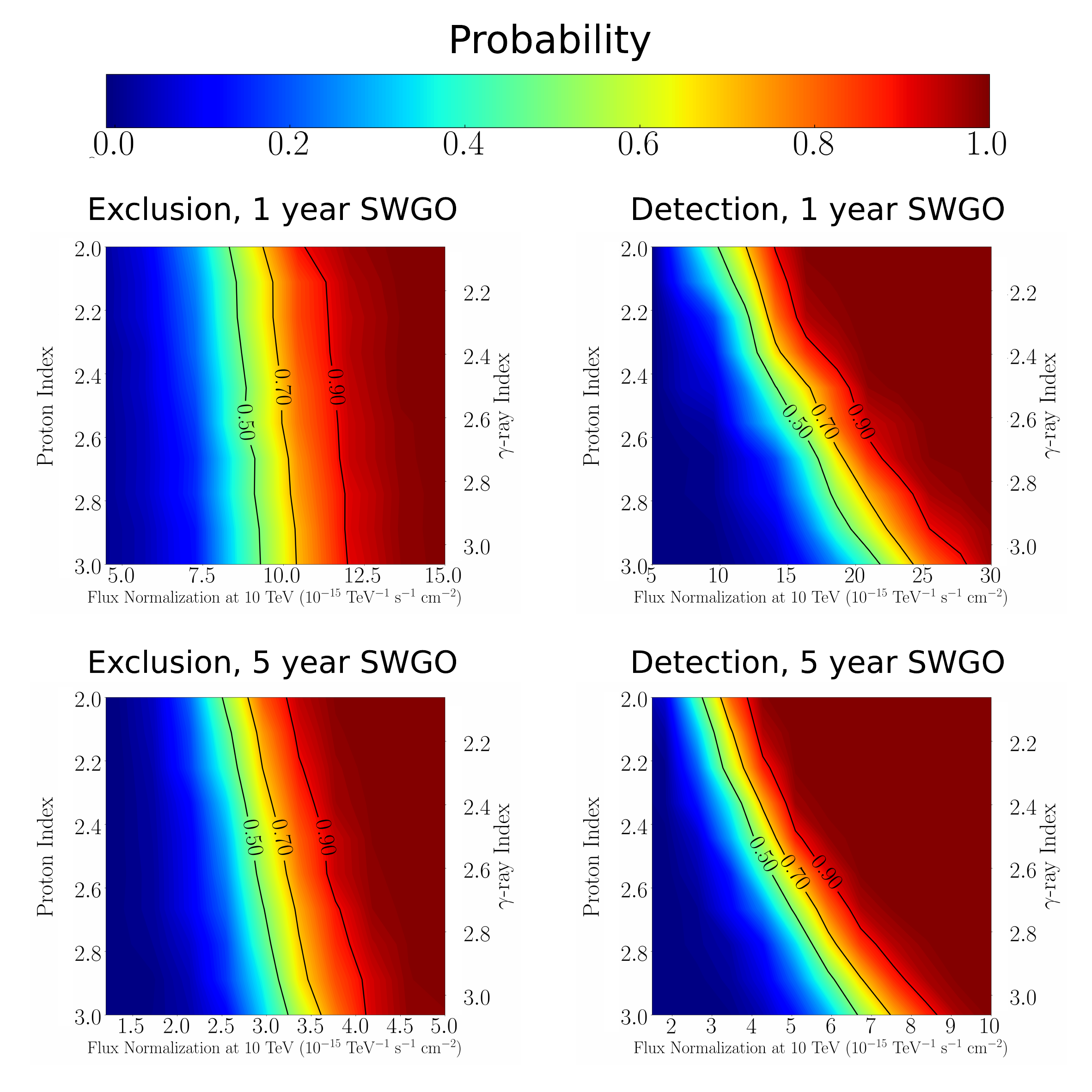}
\caption{Estimated probability maps for exclusion (left panels) and detection (right panels) of a PeVatron source with a robust statistical significance of $5\sigma$ with SWGO data. This estimation is based on the assumed SWGO straw-man sensitivity curve \citep{swgo_white} taken from https://github.com/harmscho/SGSOSensitivity. The reconstruction of maps follows the methodology introduced in \cite{cta_pevatrons}. The x-axis and the left and right y-axes represent the assumed parameter values for the true observed $\gamma$-ray flux normalization $\Phi_0$ at $10$~TeV originating from pp interactions observed from Earth, the spectral index of the hadronic particle distribution ($\Gamma_\mathrm{p}$), and the corresponding spectral index of ECPL $\gamma$-ray emission ($\Gamma_\mathrm{\gamma}$), respectively. The color bar indicates the probability of either exclusion (left panels) or detection (right panels) of a PeVatron with a statistical significance of $5\sigma$, using the PTS method. The top panels present the results for 1 year of SWGO observations, whereas the bottom panels shows 5 year observations. The cutoff energy for the hadronic particle spectrum in the left panels, representing exclusion power at a $5\sigma$ significance for non-PeVatron sources, is assumed to be $E_\mathrm{cut,p}$=$300$~TeV. On the other hand, $E_\mathrm{cut,p}$=$3$~PeV is used for the right panels, which demonstrate SWGO's power to robust detection of a PeVatron. The contours representing detection and rejection probabilities of 0.5, 0.7, and 0.9 are shown with solid black lines.}
\label{fig_ptspower}
\end{figure*}

As it was discussed in \cite{cta_pevatrons,ozi_review}, it is important to highlight that establishing a direct relationship between the 'detection or absence of $\gamma$-ray spectral cutoffs' and the 'identification of PeVatron spectral signatures' is not always straightforward. Indeed, the presence of a significant $\gamma$-ray spectral cutoff observed at UHEs (i.e. E$>$100~TeV) could potentially be interpreted as a sign of PeVatron detection assuming that the observed $\gamma$-ray emission originates from hadronic interactions. Conversely, a source that does not exhibit a clear $\gamma$-ray spectral cutoff within the energy range of the instrument, i.e.~due to its very high underlying hadronic spectral cutoff, would clearly display a more promising spectral PeVatron signature, only if this spectral behaviour is significant. The recently introduced PTS method \citep{cta_pevatrons} offers a gauge to quantitatively assess the statistical significance of such spectral behaviors.

In this section, the potential of SWGO observations to decide whether a given source is associated with a PeVatron or not is estimated based on the straw-man design configuration for general point-like $\gamma$-ray sources, assuming 1 year and 5 years of simulated SWGO observations. The investigation encompasses a wide range of parameters, denoted as $\rm{\Phi_{0}}$ which corresponds to true $\gamma$-ray flux observed from Earth, resulting from pp interactions followed by subsequent $\pi^{0}$ decay, and proton spectral index $\Gamma_\mathrm{P}$, associated with PeVatron sources. The SWGO flux points are simulated using $\gamma$-ray emission models from hypothetical PeVatrons characterized by a proton cutoff energy of $E_\mathrm{cut,p}=3$~PeV, as well as from non-PeVatron sources having a proton cutoff energy of $E_\mathrm{cut,p}=300$~TeV, taking into account various combinations of ($\rm{\Phi_{0}}$, $\Gamma_\mathrm{p}$) parameters. Since SWGO's sensitivity is expected to be more pronounced at higher energies (E$\gg$1~TeV, see Fig.~\ref{fig_fluxmodels}), the flux normalization of the maps is established at a reference energy of 10~TeV, deviating from the typical value of 1~TeV generally used in VHE astronomy. The probability to detect a PeVatron and, respectively, to exclude that a hadronic $\gamma$-ray source is a PeVatron, with a statistical significance of more than robust $S_\mathrm{PTS}$=$5\sigma$ level, is estimated by taking the fraction of simulated sources for which the PTS is larger than 25 and, respectively, smaller than $-25$ as it was discussed and detailed in \cite{cta_pevatrons,ozi_review}.

Figure~\ref{fig_ptspower} shows the SWGO PeVatron detection (right panels) and exclusion (left panels) maps, revealing that the sensitivity achieved with the SWGO straw-man design configuration is already promising for exploring significant portions of the investigated parameter space of PeVatron sources. For the purpose of relative performance comparison with future VHE experiments, such as the CTA \citep{cta_pevatrons}, similar maps reconstructed using the flux normalization at $\Phi_{0}$ at 1~TeV (in mCrab units) are also provided in \ref{appendix1}. Considering the straw-man design configuration, it becomes evident that the PeVatron detection capability of SWGO after 1 year of observations is comparable to what can be achieved with 50 hours of CTA observations. Consequently, 1-year SWGO observations can provide much higher PeVatron detection sensitivity with respect to what can be expected from the planned CTA scan of the Galactic plane \citep{cta_science, cta_gps2023}, which is estimated to have an average exposure of $\sim$10~hours. Moreover, extending observations to 5 years with SWGO yields a relatively better PeVatron detection sensitivity compared to 100 hours of deeper follow-up CTA observations. However, this comparison neglects the fact that a source which appears as point-like for SWGO, considering its angular resolution of $\sim0.15^\circ$ above $30$~TeV, will indeed not display point-like characteristics for the CTA due to its superior angular resolution. It is important to point out that the comparison based on the analysis results do not include systematic errors. The conclusions can change depending on the extent of systematic uncertainties. The power of PeVatron detection and exclusion, derived from observations with SWGO as shown in Fig.~\ref{fig_ptspower} (and also in Fig.~\ref{fig_ptspower_1TeV}), diminishes by a factor of 2$-$8, depending on the source brightness and spectral index, when a conservative systematic flux error of $\xi=10\%$ is assumed. The effect is more pronounced for the weak and soft sources, consequently shifting the transition regions\footnote{Transition region is the part of the parameter space in which the detection (or exclusion) probabilities are between 0.5 and 0.9 contour lines.} to higher flux levels. This confirms that the control of systematic errors is an important prerequisite to reach the full potential of SWGO.

\section{Probing promising PeVatron candidates of the Southern-sky with SWGO }
\label{promising_sources}
\begin{figure*}[!ht]
\centering
\includegraphics[width=18cm]{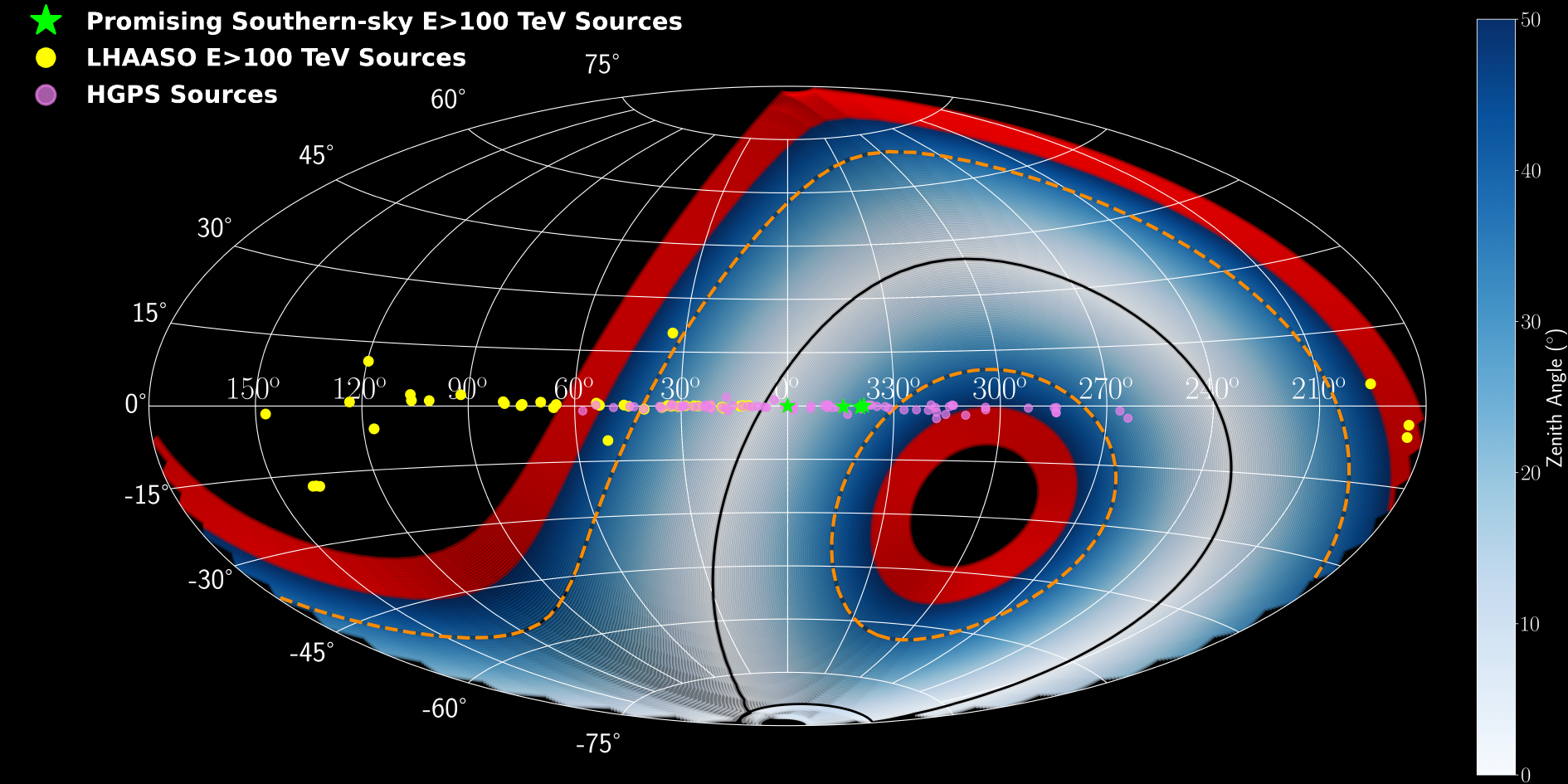}
\caption{
The SWGO visibility plot for the Southern-sky assuming the observatory situated at 23$^{\circ}$ South latitude, which is a similar latitude location of the H.E.S.S. experiment \citep{hess_crab}. The gradient of blue colors illustrates the visibility of objects at different observation zenith angles, while the red shaded areas indicate sky regions where the observation zenith angle is between 40$^{\circ}$ and 50$^{\circ}$, which are generally considered as sub-optimal observation conditions for obtaining reliable data. The black solid and dashed orange lines correspond to the sky visibility at the zenith angles of 0$^{\circ}$ and 30$^{\circ}$, respectively. The yellow and magenta dots highlighted on the plot mark the position of the LHAASO sources showing significant (S$_\mathrm{UHE}$ $\geq$ 4$\sigma$) E$>$100~TeV $\gamma$-ray emission regions \citep{lhaaso_1stcat} and VHE sources detected in the H.E.S.S. Galactic Plane Survey \citep{hgps}, respectively. The green stars mark the location of promising E$>$100~TeV Southern sky sources, namely the GC diffuse emission region, HESS~J1702$-$420A, Westerlund~1 and HESS~J1641$-$463, discussed extensively in Sec.~\ref{promising_sources}. The plot is produced using the {\tt swgo-plot} module provided in 
https://swgo-collaboration.gitlab.io/swgo-plot/.} 
\label{fig_swgo_visibility}
\end{figure*}

\begin{table*}
\caption{Spectral characteristics of the promising Southern-sky E$>$100~TeV $\gamma$-ray sources, the GC diffuse emission region \citep{hess_gc_pevatron}, HESS~J1702$-$420A \citep{j1702}, Westerlund~1 \citep{westerlund1} and HESS~J1641$-$463 \citep{j1641}. The differential flux at 1~TeV is provided in both TeV$^{-1}$cm$^{-1}$s$^{-1}$ and mCrab units in the second and third columns, while the corresponding differential flux at 10~TeV is given in the fourth column. Details regarding the best-fit $\gamma$-ray spectral indices, spectral cutoffs (if detected), and the favored $\gamma$-ray models are provided in the fifth, sixth, and seventh columns, respectively. 
We note that the $\Phi_{0}$(1~TeV) value of HESS~J1702$-$420A is derived by extrapolating the $\gamma$-ray model described in \cite{j1702} down to 1~TeV energy. The E$_\mathrm{Max}$ column gives the energy of the highest significant flux point, while the energy of the highest flux upper limits (if exists) are provided in parentheses.}
\label{source_prop}     
\centering       
\renewcommand{\arraystretch}{1.4}
\begin{tabular}{c c c c c c c c c}       
\hline\hline                 
Source  & $\Phi_{0}$(1~TeV) & $\Phi_{0}$(1~TeV) & $\Phi_{0}$(10~TeV) & Spectral   & E$_\mathrm{cut,\gamma}$ & Preferred & E$_\mathrm{Max}$ \\
Name & (TeV$^{-1}$cm$^{-1}$s$^{-1}$) & (mCrab) & (TeV$^{-1}$cm$^{-1}$s$^{-1}$) & Index & (TeV) & Model & (TeV) \\ 
\hline       
GC diffuse emission         & (1.92$\pm$0.08)$\times$10$^{-12}$ & $\sim$50 & $\sim$9.2$\times$10$^{-15}$  & 2.32$\pm$0.05 & $-$ & PL & 39.6 (58.8)\\
HESS J1702$-$420A & $\sim$1.6$\times$10$^{-13}$ & $\sim$4 &  $\sim$4.7$\times$10$^{-15}$ & 1.53$\pm$0.19 & $-$ & PL & 84.8 (130.1) \\
Westerlund 1      & (1.00$\pm$0.03)$\times$10$^{-11}$ & $\sim$260 & $\sim$4.0$\times$10$^{-14}$ & 2.30$\pm$0.04 & 44$^{+17}_{-11}$ & ECPL & 80.6 \\
HESS J1641$-$463  & (3.91$\pm$0.69)$\times$10$^{-13}$ & $\sim$10 & $\sim$3.3$\times$10$^{-15}$ & 2.07$\pm$0.11 & $-$ & PL & 23.4 (68.7)  \\
\hline  
\hline   
\end{tabular}
\end{table*}

In this section, the analysis results of SWGO simulations based on the public spectral $\gamma$-ray data from four promising Southern-sky Galactic PeVatron candidate sources, the GC~diffuse emission\footnote{The GC diffuse emission spectrum is extracted from an annulus centred at Sgr A* (see right panel of Fig.~1 in \cite{hess_gc_pevatron}) with inner and outer radius of 0.15$^{\circ}$ and 0.45$^{\circ}$, respectively, and a solid angle of 1.4$\times$10$^{-4}$ sr.} region at the center of the Galaxy \citep{hess_gc_pevatron}, the young massive stellar cluster Westerlund~1 \citep{westerlund1}, and unidentified hard $\gamma$-ray sources HESS~J1702$-$420A \citep{j1702} and HESS~J1641$-$463 \citep{j1641}, are presented and discussed in the framework of the SWGO straw-man configuration design. All four of these $\gamma$-ray sources detected at VHEs are considered as promising E$>$100~TeV emitters, either due to their fluxes detected at high-energies (such as $\sim$81~TeV for Westerlund~1 and $\sim$85~TeV for HESS~J1702$-$420A) or from their observed power-law characteristic that exhibit hard spectral features without showing any clear indications of spectral cutoffs (GC~diffuse emission region and HESS~J1641$-$463). Such unique spectral features, which are listed in Table~\ref{source_prop}, make these sources particularly intriguing as potential Southern-sky PeVatron candidates, as the possibility of hadronic emission scenarios cannot be ruled out for any of them. Location of these Southern-sky PeVatron candidates are marked with green dots in the SWGO Southern-sky visibility plot shown Fig.~\ref{fig_swgo_visibility}. As illustrated, all of these sources can be observed under ideal conditions, with zenith angles smaller than 30$^{\circ}$, assuming the observatory being situated $\sim$23$^{\circ}$ South latitudes. These promising candidates discussed in this section are presumed to be spatially isolated and are treated as point-like sources for SWGO. The assumption of a point-like source holds true for HESS~J1702$-$420A and HESS~J1641$-$463, since their spatial extensions of $0.06^\circ\pm 0.02^\circ\,(\mathrm{stat})\,\pm 0.03^\circ\,(\mathrm{sys})$ \citep{j1702} and an upper limit of $0.05^\circ$ \citep{j1641} degrees, respectively, are significantly smaller than SWGO's design angular resolution of $\sim$0.15$^{\circ}$ above 30~TeV. However, the GC diffuse emission region is defined with an outer radius of $0.45^\circ$ \citep{hess_gc_pevatron}, and the emission region around Westerlund~1 extends up to a diameter of $\sim$2$^{\circ}$ \citep{westerlund1}, clearly deviating from the point-like source assumption for SWGO. Additionally, it's important to note that all sources are actually not "spatially isolated" and they potentially suffer from the effects of source confusion, where the presence of multiple nearby $\gamma$-ray sources complicates the analysis in practice. We explicitly state that due to these idealized assumptions, the following estimates can only serve as first benchmarks within simplified conditions.

\subsection{The GC diffuse emission region and the unidentified source HESS~J1702$-$420A}

The PTS analysis of the available VHE $\gamma$-ray data from the H.E.S.S. experiment for the GC diffuse emission and HESS~J1702$-$420A regions was presented and extensively discussed 
in \cite{pts_paper}, resulting in non-significant $\mathrm{S}_\mathrm{PTS}$ values of 0.4$\sigma$ and 1.0$\sigma$, respectively.~Assuming $\xi=20\%$ systematic error in the H.E.S.S. data sets, the corresponding $95\%$~CL cutoff lower limits were calculated for the underlying hadronic spectra, yielding 172~TeV for the GC diffuse emission region and 436~TeV for HESS~J1702$-$420A. Due to these results, it was not possible to draw any statistically significant conclusions regarding the PeVatron nature of these sources. However, it was discussed that the potential impact of forthcoming observatories, particularly the SWGO experiment, is crucial since both sources lack UHE data above 100~TeV, and data from such future observatories could provide crucial insights on determining whether these sources can be classified as Galactic PeVatrons.

\subsection{Westerlund~1: The young massive stellar cluster}

One of the regions in the Southern-sky showing potential promise for emitting UHE $\gamma$-rays above 100~TeV is the Westerlund~1 region, which stands out as the most massive young stellar cluster (YMC) within our Galaxy, estimated to be around 4$-$5 Myr old according to \cite{wd1_age}. The YMC environments in our Galaxy serve as fertile grounds for star formation, containing stars in the early stages of formation, surrounded by cosmic gas and dust. As a result of such conditions, these environments become efficient regions for accelerating particles which can interact with the surrounding gas, and are considered as promising Galactic PeVatron candidates \citep{morlino_sc}. A recent detailed VHE analysis of the region using H.E.S.S. data revealed the presence of an extended and complex shell-like $\gamma$-ray emission spanning up a diameter of $\sim$2$^{\circ}$. The authors mentioned that even though there is not a clear spatial correlation between the structures of interstellar gas and the observed VHE $\gamma$-ray emission, the possibility of a scenario involving hadronic interactions is still possible due to the lack of energy-dependent morphology and uncertainties of the gas distribution \citep{westerlund1}.

The ECPL model, given in Eq.~\ref{eq1}, was fitted to the $\gamma$-ray flux data points of Westerlund~1, considering systematic errors of $\xi=20\%$, resulting in $\Phi_{0}$(1~TeV)=(1.02$\pm$0.07)$\times$10$^{-11}$~TeV$^{-1}$cm$^{-1}$s$^{-1}$, $\Gamma_{\gamma}$=2.33$\pm$0.08 and E$_\mathrm{cut,\gamma}$=44.2$\pm$21.4 TeV, which are consistent with the findings presented in \cite{westerlund1}. In this context, the statistical significance of the $\gamma$-ray spectral cutoff feature is found to be $\mathrm{S}_\mathrm{cut,\gamma}$=2.4$\sigma$, and the corresponding lower limit for the $\gamma$-ray cutoff at the $95\%$~CL~is LL$_\mathrm{cut,\gamma}$=23~TeV. Assuming that the entire $\gamma$-ray emission arises from interactions between accelerated protons, following the spectral shape defined in Eq.~\ref{eq2}, and target gas in the region, the spectral index and cutoff parameters of the underlying parental proton spectral can  be obtained as $\Gamma_\mathrm{p}$=2.33$\pm$0.12 and $E_\mathrm{cut,p}$=300$\pm$188~TeV, respectively. The statistical significance of the proton cutoff feature is calculated as $\mathrm{S}_\mathrm{cut,p}$=2.4$\sigma$, while the $95\%$~CL lower limit for the proton cutoff is derived at 127~TeV. The corresponding PTS significance for the overall emission originating from the Westerlund 1 region is estimated to be $\mathrm{S}_\mathrm{PTS}$=$-$1.4$\sigma$. When considering only the available H.E.S.S. data, a conclusive determination cannot be made regarding whether the emission from Westerlund~1 arises from PeVatron activity. Consequently, further observations at UHE, particularly above 100~TeV, are crucial to reach a conclusive determination of the PeVatron nature of Westerlund~1.

\subsection{The unidentified source HESS~J1641$-$463}
The unidentified source HESS~J1641$-$463 in the Southern-sky presents another promising region for the emission of E$>$100~TeV $\gamma$-rays. The source exhibits a hard VHE $\gamma$-ray spectrum extending up to a few tens of TeV without showing any significant spectral cutoff \citep{j1641, hess1641_proc}. Similar to the situation with HESS~J1702$-$420A, the source is affected by source confusion due to the presence of the bright and extended ($\sim$0.11$^{\circ}$) $\gamma$-ray source HESS~J1640$-$465 \citep{hess1640_paper}, which is located $\sim$0.28$^{\circ}$ away and showing a significant $\gamma$-ray cutoff in its spectrum at $\sim$6~TeV. There are dense molecular clouds found toward the line of sight, along with two nearby supernova remnants, SNR~G338.3$-$0.0 and SNR~G338.5$+$0, while the latter is found to be spatially coincident with the source. These neighboring SNRs could potentially serve as sources of accelerated protons, suggesting a plausible hadronic scenario for the observed emission. By using the $\gamma$-ray flux data points from HESS~J1641$-$463 and assuming a $\xi=20\%$ systematic error, lower limits for the $\gamma$-ray spectral cutoff at a 95$\%$ CL are derived as LL$_\mathrm{cut,\gamma}$=12.4~TeV. Under the hypothesis that the entire $\gamma$-ray emissions originates from a hadronic interactions, the spectral index of the parent protons and the 95$\%$ CL lower limit for the proton cutoff parameters can be determined as $\Gamma_\mathrm{p}$=2.03$\pm$0.15 and LL$_\mathrm{cut,p}$=64.1~TeV, respectively, while the significance of the PTS is found to be $\mathrm{S}_\mathrm{PTS}$=0.6$\sigma$. Similarly to the Westerlund~1 case, robust conclusions on the PeVatron nature of HESS~J1641$-$463 cannot be drawn when taking into account only the H.E.S.S. data, emphasizing the need for additional UHE data to provide further insights.

\subsection{SWGO simulations and data analysis of the promising Southern-sky PeVatron candidates}

\begin{figure*}[!ht]
\centering
\includegraphics[width=18cm]{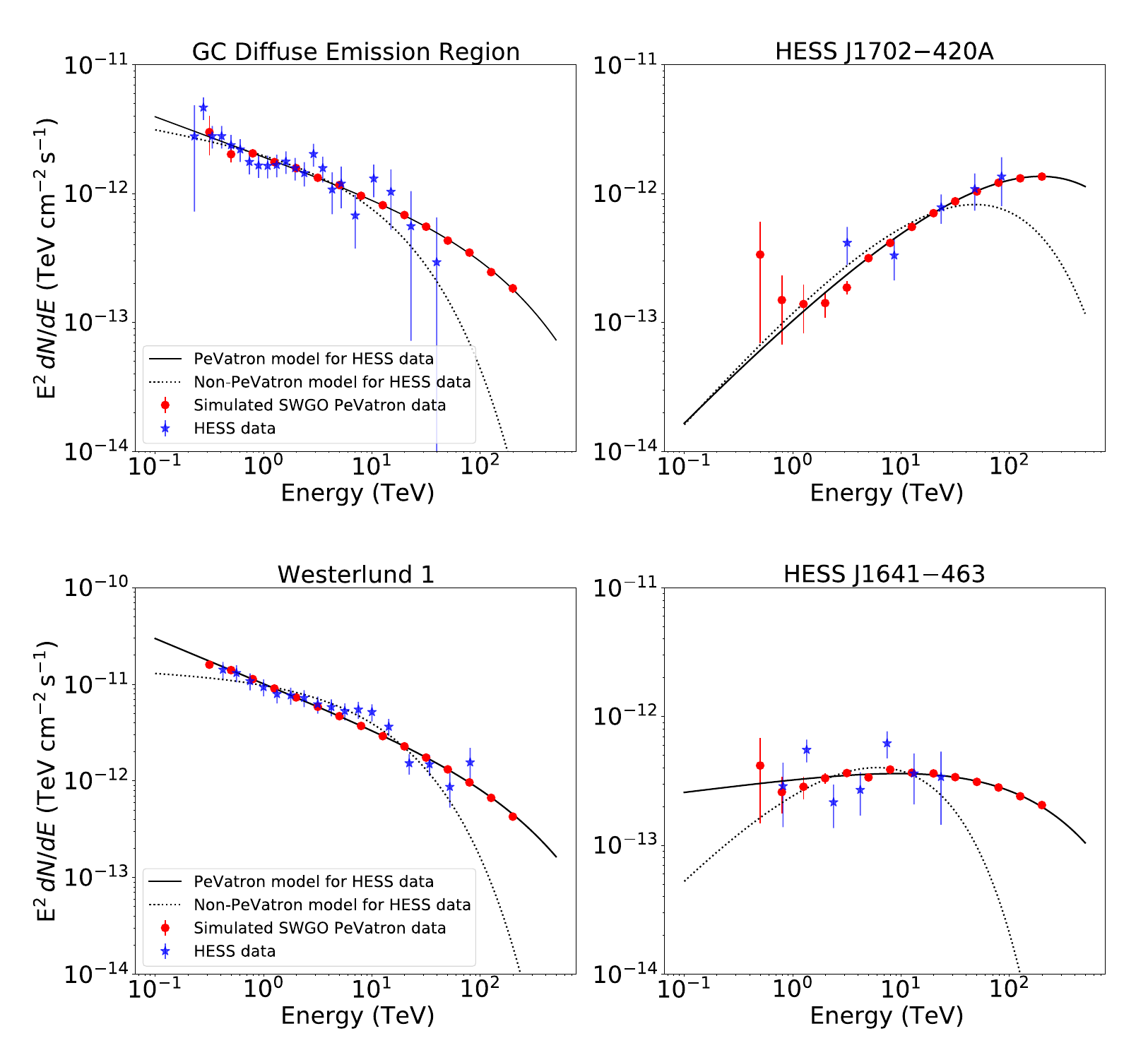}
\caption{Simulated $\gamma$-ray spectra from 5~years of SWGO data for the GC diffuse emission region~(upper left panel), HESS~J1702-420A~(upper right panel), Westerlund~1~(lower left panel) and HESS~J1641$-$463 (lower right panel). The 5-year straw-man configuration sensitivity of SWGO taken from \text{https://github.com/harmscho/SGSOSensitivity} is used in simulations. Spectral $\gamma$-ray flux data from observations with H.E.S.S. are shown in blue stars with an assumed minimal relative flux error of $\xi=20\%$. Solid black lines show the best fit PeVatron models with a hadronic energy cutoff fixed at $3$~PeV, while the dashed black lines are the best-fit non-PeVatron models in which hadronic energy cutoffs are fixed the respective $95\%$~CL lower limit derived from the H.E.S.S. data. The SWGO flux points obtained from the simulation of the best fit PeVatron models are shown in red without considering any systematic errors ($\xi=0\%$).} 
\label{fig_swgo_ex}
\end{figure*}

For all the sources discussed in this section, the available spectral H.E.S.S. data assuming a minimum relative flux error of $\xi=20\%$ are fitted to the hadronic emission model defined by Eq.~\ref{eq2}. The resulting best-fit hadronic models obtained from the analysis of H.E.S.S. data are adjusted and further used in the SWGO simulations of each individual source under investigation. An example simulation result of 5 years SWGO observations is shown in Fig.~\ref{fig_swgo_ex} for the diffuse emission region in the vicinity of the GC (top left panel), the unidentified $\gamma$-ray source HESS~J1702$-$420A (top right panel), Westerlund~1 region (bottom left panel) and HESS~J1641$-$463 (bottom right panel). The solid lines in Fig.~\ref{fig_swgo_ex} represent the hadronic models reconstructed from H.E.S.S. data with an energy cutoff fixed at 3~PeV. These hypothetical PeVatron models correspond to the expected $\gamma$-ray emission from Galactic PeVatrons, which can significantly contribute to the knee structure observed in the CR spectrum. On the other hand, the dashed lines in the figure represent best-fit non-PeVatron models derived from the same H.E.S.S. data. For these non-PeVatron models, the hadronic energy cutoff is fixed to the derived $95\%$~CL~lower limit of the underlying proton spectral cutoff obtained from respective H.E.S.S. observations: 172~TeV for the GC diffuse emission region, 436~TeV for HESS~J1702$-$420A, 127~TeV for Westerlund~1, and 64~TeV for HESS~J1641$-$463. These non-PeVatron models, characterized by proton cutoff energies fixed at their respective $95\%$~CL lower limits, remain significantly below 3~PeV, consequently are not expected to substantially contribute to the knee feature. To assess the impact of systematic errors, the SWGO flux data for PeVatron and non-PeVatron models are simulated considering statistical errors only ($\xi=0$), and with an additional optimistic $\xi=5\%$ and conservative $\xi=10\%$ systematic errors. The red SWGO flux data points shown in Fig.~\ref{fig_swgo_ex} account only for statistical errors ($\xi=0$).

\begin{table*}[ht!]
\caption{Analysis results inferred from the SWGO simulations of four promising Southern-sky E$>$100~TeV sources. Each source instance is simulated 500 times and the results are obtained from the respective distributions of derived properties. The 'PeV' and 'Non-PeV' identifiers next to source names indicate whether the PeVatron (E$_\mathrm{cut,\,p}$ fixed to 3~PeV) or Non-PeVatron (E$_\mathrm{cut,\,p}$ fixed to derived 95$\%$ proton cutoff lower limits) hadronic models used when simulating the SWGO flux data, while $\xi$ is the assumed minimal relative flux error explained in Sec.~\ref{sim_analyze}. S$_\mathrm{PTS}$ and LL$_\mathrm{cut,\,p}$ denote the significance of the PTS and the $95\%$ CL lower limit on the hadronic energy cutoff E$_\mathrm{cut,\,p}$, while $\mathrm{S}_\mathrm{cut,p}$ is the significance of the hadronic cutoff feature, respectively. Similarly, LL$_\mathrm{cut,\gamma}$ and $\mathrm{S}_\mathrm{cut,\gamma}$ are the corresponding $95\%$ CL lower limit on the $\gamma$-ray cutoff energy E$_\mathrm{cut,\,p}$ and the significance of the $\gamma$-ray cutoff feature. The derived reference properties, on which conclusions are based in this work, are highlighted in bold and shown in Fig.~\ref{fig_swgo_ex}.}
\label{individual_sources_table}
\centering
\renewcommand{\arraystretch}{1.4}
\begin{tabular}{llllllllll}
\\
\hline\hline
Source Name & $\xi$  & $\mathrm{S}_\mathrm{PTS}$ & LL$_\mathrm{cut,p}$ & $\mathrm{S}_\mathrm{cut,p}$ & E$_\mathrm{cut,p}$ & LL$_\mathrm{cut,\gamma}$ & $\mathrm{S}_\mathrm{cut,\gamma}$ & E$_\mathrm{cut,\gamma}$\\
            & ($\%$) & ($\sigma$)  & (TeV) & ($\sigma$) & (TeV) & (TeV) & ($\sigma$) & (TeV)\\
\hline\hline
\textbf{GC diffuse emission (PeV)}     & \textbf{0}  & \textbf{11.7$\pm$1.0}    & \textbf{2479$\pm$285} & \textbf{11.9$\pm$1.0} & \textbf{3020$\pm$391} & \textbf{244$\pm$21} & \textbf{11.5$\pm$1.0} & \textbf{282$\pm$28}\\
GC diffuse emission (PeV)     & 5 & 6.6$\pm$0.7     & 2153$\pm$314 & 6.6$\pm$0.8  & 3022$\pm$545 & 240$\pm$27 & 6.4$\pm$0.8 & 307$\pm$43\\
GC diffuse emission (PeV)     & 10 & 4.1$\pm$0.6     & 1795$\pm$312 & 4.2$\pm$0.7  & 2969$\pm$759 & 220$\pm$26 & 4.1$\pm$0.8 & 316$\pm$58\\
\hline
GC diffuse emission (Non-PeV) & 0  & $-$22.4$\pm$1.0 & 158$\pm$9    & 33.8$\pm$1.1 & 173$\pm$10 & 33$\pm$1 & 33.6$\pm$1.1 & 36$\pm$1\\
GC diffuse emission (Non-PeV) & 5  & $-$12.8$\pm$0.7 & 148$\pm$9    & 18.8$\pm$0.7 & 172$\pm$11 & 34$\pm$2 & 18.6$\pm$0.7 & 37$\pm$2\\
GC diffuse emission (Non-PeV) & 10  & $-$8.6$\pm$0.7 & 139$\pm$13    & 12.3$\pm$0.7 & 173$\pm$17 & 34$\pm$3 & 12.1$\pm$0.7 & 40$\pm$3\\
\hline\hline
\textbf{HESS J1702$-$420A (PeV)}     & \textbf{0}  & \textbf{19.8$\pm$1.2}    & \textbf{2593$\pm$243} & \textbf{25.2$\pm$0.9} & \textbf{3053$\pm$290}  & \textbf{271$\pm$10}  & \textbf{25.0$\pm$0.9} & \textbf{290$\pm$12}   \\ 
HESS J1702$-$420A (PeV)     & 5 & 4.3$\pm$0.3     & 1773$\pm$329  & 7.2$\pm$0.5  & 3021$\pm$592 & 201$\pm$11  & 7.3$\pm$0.4  & 248$\pm$19  \\ 
HESS J1702$-$420A (PeV)     & 10 & 2.3$\pm$0.3     & 1250$\pm$332  & 4.1$\pm$0.5  & 3100$\pm$1012 & 166$\pm$12  & 4.2$\pm$0.4  & 236$\pm$24  \\ 
\hline
HESS J1702$-$420A (Non-PeV) & 0  & $-$11.0$\pm$1.0 & 427$\pm$7    & 45.5$\pm$1.0 & 441$\pm$21    & 102$\pm$3   & 44.0$\pm$1.0 & 107$\pm$3  \\ 
HESS J1702$-$420A (Non-PeV) & 5 & $-$4.9$\pm$0.7  & 415$\pm$9   & 18.2$\pm$0.7  & 449$\pm$32    & 97$\pm$3    & 17.7$\pm$0.7  & 107$\pm$4  \\ 
HESS J1702$-$420A (Non-PeV) & 10 & $-$2.8$\pm$0.5  & 397$\pm$10   & 10.2$\pm$0.5  & 453$\pm$48    & 86$\pm$4    & 10.0$\pm$0.4  & 102$\pm$5  \\
\hline\hline
\textbf{Westerlund 1 (PeV)}     & \textbf{0}  & \textbf{37.5$\pm$1.0}     & \textbf{2805$\pm$109} & \textbf{34.6$\pm$1.0}  & \textbf{2996$\pm$121}  & \textbf{279$\pm$8} & \textbf{34.3$\pm$1.0} & \textbf{294$\pm$9}\\
Westerlund 1 (PeV)     & 5 & 8.5$\pm$0.5      & 2290$\pm$198 & 8.1$\pm$0.7   & 2899$\pm$313 & 281$\pm$18 & 7.9$\pm$0.7  & 344$\pm$27 \\
Westerlund 1 (PeV)     & 10 & 4.3$\pm$0.2      & 1832$\pm$126 & 4.1$\pm$0.3   & 2999$\pm$296 & 237$\pm$12 & 4.0$\pm$0.3  & 342$\pm$26 \\
\hline
Westerlund 1 (Non-PeV) & 0  & $-$141.3$\pm$1.0 & 126$\pm$1   & 197.7$\pm$1.3 & 128$\pm$1    & 28$\pm$1 & 196.5$\pm$1.3 & 29$\pm$1\\
Westerlund 1 (Non-PeV) & 5  & $-$26.5$\pm$0.6 & 118$\pm$3   & 34.8$\pm$0.7 & 128$\pm$4    & 31$\pm$1 & 34.3$\pm$0.8 & 33$\pm$1\\
Westerlund 1 (Non-PeV) & 10 & $-$16.2$\pm$0.6 & 113$\pm$6   & 20.5$\pm$0.6 & 128$\pm$7    & 32$\pm$2 & 20.1$\pm$0.6 & 36$\pm$2\\
\hline\hline
\textbf{HESS J1641$-$463 (PeV)}     & \textbf{0}  & \textbf{6.7$\pm$1.0}     & \textbf{2225$\pm$398} & \textbf{8.9$\pm$0.1}  & \textbf{3050$\pm$625}  & \textbf{219$\pm$25} & \textbf{8.3$\pm$1.0}  & \textbf{268$\pm$37} \\
HESS J1641$-$463 (PeV) & 5 & 5.2$\pm$0.8  & 2032$\pm$392   & 6.5$\pm$0.8  & 3045$\pm$717    & 211$\pm$25    & 6.1$\pm$0.8  & 273$\pm$41  \\
HESS J1641$-$463 (PeV) & 10 & 3.6$\pm$0.6  & 1721$\pm$368   & 4.3$\pm$0.7  & 3065$\pm$930    & 191$\pm$23    & 4.0$\pm$0.7  & 275$\pm$50  \\
\hline
HESS J1641$-$463 (Non-PeV) & 0  & $-$18.5$\pm$1.0 & 56$\pm$4     & 24.7$\pm$1.1 &     65$\pm$7      & 21$\pm$1   & 24.5$\pm$1.2 & 23$\pm$1\\
HESS J1641$-$463 (Non-PeV) & 5 & $-$14.8$\pm$0.9  & 55$\pm$4   & 19.2$\pm$0.9  & 65$\pm$8    & 21$\pm$2    & 19.1$\pm$0.9  & 23$\pm$2  \\
HESS J1641$-$463 (Non-PeV) & 10 & $-$10.0$\pm$0.7  & 54$\pm$4   & 12.9$\pm$0.7  & 66$\pm$10    & 20$\pm$2    & 12.7$\pm$0.7  & 23$\pm$3  \\

\hline\hline
\end{tabular}
\end{table*}

In order to accumulate reliable statistics, the simulation procedure described above is repeated 500 times for all examined sources, and $S_\mathrm{PTS}$ values are calculated assuming PeVatron and non-PeVatron models considering systematic errors of $\xi=0$, $\xi=5\%$ and $\xi=10\%$ in each simulated SWGO data set. In addition, the best fit proton and $\gamma$-ray spectral cutoff energies (E$_\mathrm{cut,p}$ and E$_\mathrm{cut,\gamma}$), statistical significance of the respective proton and $\gamma$-ray cutoff features ($\mathrm{S}_\mathrm{cut,p}$ and $\mathrm{S}_\mathrm{cut,\gamma}$) and the 95$\%$ CL of proton and $\gamma$-ray spectral cutoff lower limits (LL$_\mathrm{cut,p}$ and LL$_\mathrm{cut,\gamma}$) are derived for each simulated SWGO data set. These characteristics derived from 500 simulations are then gathered into distributions, and their median values along with standard errors are computed and summarized in Table~\ref{individual_sources_table}.

The derived intrinsic properties mentioned above have the potential to provide insights about the PeVatron characteristics of sources. The results of the simulations clearly indicate that if the examined $\gamma$-ray sources are associated to hadronic Galactic PeVatrons, which contribute to the knee feature observed at 3~PeV energies, SWGO possesses substantial potential to confirm their PeVatron nature at a robust CL ($\mathrm{S}_\mathrm{PTS}$~$\geq$~5.0$\sigma$). Similarly, the absence of PeVatron characteristics can also be robustly confirmed ($\mathrm{S}_\mathrm{PTS}$~$\leq$~$-$5.0$\sigma$) for all sources, provided that the corresponding proton energy cutoffs are well below 3~PeV. Given the specific cutoff values assigned to the assumed respective proton spectra, the application of the PTS technique using spectral data inferred from SWGO observations allow a robust determination of whether the $\gamma$-ray emissions from these promising Southern-sky E$>$100~TeV sources are Galactic PeVatrons in nature. 

When a conservative $\xi=10\%$ SWGO systematic error is considered alongside with otherwise unchanged simulation parameters, the median significances of the PTS values are diminished to a marginal detection range of $\mathrm{S}_\mathrm{PTS}\cong$3$-$4$\sigma$ for PeVatron cases, while the instances where non-PeVatron characteristics can still be robustly confirmed, with the exception of HESS~J1702$-$420A. On the other hand, in the case of $\xi=5\%$ SWGO systematic error, robust detection of both PeVatron and non-PeVatron characteristics can be confirmed. These simulation results clearly highlight the importance of inferred systematics, and show that the full potential of SWGO in PeVatron searches can only be achieved when systematic errors are carefully controlled and minimized as effectively as possible. Especially, systematic flux error levels similar to LHAASO experiment or better ($\xi= 5-7\%$) can lead to significant detection of spectral PeVatron characteristics.
Furthermore, assuming point-like source morphology for the extended GC diffuse emission and Westerlund~1 regions can have significant impact on the results. As it was shown in \citep{cta_pevatrons}, both the PeVatron detection and rejection probabilities decrease as the source extension increases (see \cite{ambrogi_cta} for a detailed discussion on extended source sensitivities). Consequently, the $\mathrm{S}_\mathrm{PTS}$ values (and lower limits, LL$_\mathrm{cut,p}$, LL$_\mathrm{cut,\gamma}$) obtained for these extended sources tend to be overestimated.

\section{Discussions and Conclusions}
\label{discussions}

In practical applications, it is expected that the issue of source confusion, which refers to the condition where multiple $\gamma$-ray sources exist within an unresolved spatial distance, becomes a significant challenge when analysing Galactic $\gamma$-ray data. This challenge is particularly pronounced when dealing with $\gamma$-ray energies below 10~TeV, given that many Galactic VHE sources are known either to exhibit cutoffs in their spectra below these energies, or not to emit significant flux above 10~TeV due to soft power-law spectral index of the emission. As a result, the problem of source confusion becomes more relevant with decreasing $\gamma$-ray energy. One particular example is the unidentified source HESS~J1702$-$420 discussed in this paper. The investigation of data from HESS~J1702$-$420 has revealed two closely positioned sub-components without the detection of significant energy cutoff \citep{j1702}. In this particular case, the component HESS~J1702$-$420A becomes more luminous than the second component, HESS~J1702$-$420B, above a few tens of TeV due to relative difference in their power-law spectral indices. Another example comes from the observations of a specific Galactic region encompassing the Boomerang~PWN and SNR~G106.3$+$2.7. The observations of this region conducted by MAGIC provided compelling evidence that supports the existence of two distinct power-law source components \citep{magic_tail}. The softer of the two components, referred to as the 'head', and the harder one, known as the 'tail', both show no clear spectral cutoffs. Moreover, the observations carried out by LHAASO in the same region \citep{lhaaso2021} exclusively detect a single source component emitting at UHEs, and as it was discussed in \cite{pts_paper}, the UHE emission detected by LHAASO can be connected to the tail emission detected by MAGIC. In contrary, when examining the case of HESS~J1641$-$463, which exhibits a hard power-law spectrum, the neighboring source HESS~J1640$-$465 shows a significant spectral cutoff in its spectrum below 10~TeV. These examples show that $\gamma$-ray emission beyond several tens of TeV can be well dominated by a single source even in the case of source confusion, particularly either when no spectral cutoff is detected or the spectral index of one component is significantly harder than for the neighboring sources. Consequently, an unprecedented level of information on comprehensive understanding of the PeVatron nature of sources can be acquired from the synergy between the experiments like CTA and SWGO due to their complementary capabilities, with CTA excelling in superior angular resolution and SWGO enhancing flux sensitivity at high energies.

As evident from both Fig.~\ref{fig_swgo_sens} and Fig.~\ref{fig_fluxmodels}, the flux sensitivity of SWGO will offer a simple detection $\gamma$-ray sources at high energies, providing an unprecedented level of statistical information beyond 10~TeV, at which concerns related to source confusion is effectively eliminated. On the other hand, the superior angular resolution of CTA will provide crucial information for resolving individual sources below a few tens of TeV energies, at which source confusion is expected to have more significant impact, and therefore supplying key insights into the source component with which the observed UHE emission can be associated. Such a clear association between a $\gamma$-ray source exhibiting distinct PeVatron spectral signatures and any type of Galactic CR accelerator is essential for robust determination of Galactic objects which are truly the PeVatrons responsible for the knee feature observed in the CR spectrum. As a result, such a synergy can potentially shed light on the century-old enigma of the origin of Galactic CRs. However, when addressing the issue of source confusion through a combined analysis of data from different observatories, such as future CTA South and SWGO, the relative systematic errors between the flux measurements of different observatories must be carefully controlled. 

In this paper, the expected potential of SWGO in PeVatron searches are investigated using the straw-man design sensitivity curve. It was concluded that the high energy $\gamma$-ray cutoffs between 30~TeV and 100~TeV can be significantly detected for relatively faint 5~mCrab sources, when the spectral index is hard ($\Gamma\lesssim$ 2.0), while the detection of spectral cutoffs for the relatively soft $\Gamma\cong$2.3 sources can only be possible if they are bright enough, i.e.~$\Phi_{0}$ $\geq$ 11~mCrab. The reconstructed SWGO PeVatron detection maps show that the SWGO can probe large parts of the investigated PeVatron parameter space, providing a robust detection and/or rejection power. A dedicated study on the promising Southern-sky E$>$100~TeV sources gives similar results, concluding that the SWGO will have a great potential to confirm or exclude their PeVatron nature at a robust significance level after 5-years of observation. The study also demonstrates that the control of SWGO systematic errors will be a necessary issue, and they should be around $5-7\%$ in order to reach the maximized potential of detecting spectral PeVatron characteristics. We explicitly mention that the results presented in this paper do not reflect a fiducial performance of the planned SWGO observatory, instead can provide a preliminary insight on the performance of PeVatron searches with SWGO. The results presented in this paper are based on straw-man design configuration, therefore they are conservative. Indeed, with the low and high energy enhancements, together with improved PSF and background rejection, the performance capabilities, especially SWGO abilities to detect spectral cutoffs at high energies and PeVatron signatures will be significantly improved.

\section*{Acknowledgements}
E.O.A. acknowledges financial support by TÜBİTAK Research Institute for Fundamental Sciences.\\
We express our sincere gratitude to Gerrit Spengler for his extremely useful active contributions and constructive feedback. We also express our sincere gratitude to the SWGO Collaboration, especially the feedbacks provided by Ulisses Barres, Ruben Concei\c{c}\~ao and Sidharth Sreeja Sadanandan, which greatly enhanced the quality of the paper.\\

\appendix
\section{Parameter scan at 1 TeV flux normalization}
\label{appendix1}

\begin{figure*}[!ht]
\centering
\includegraphics[width=17cm]{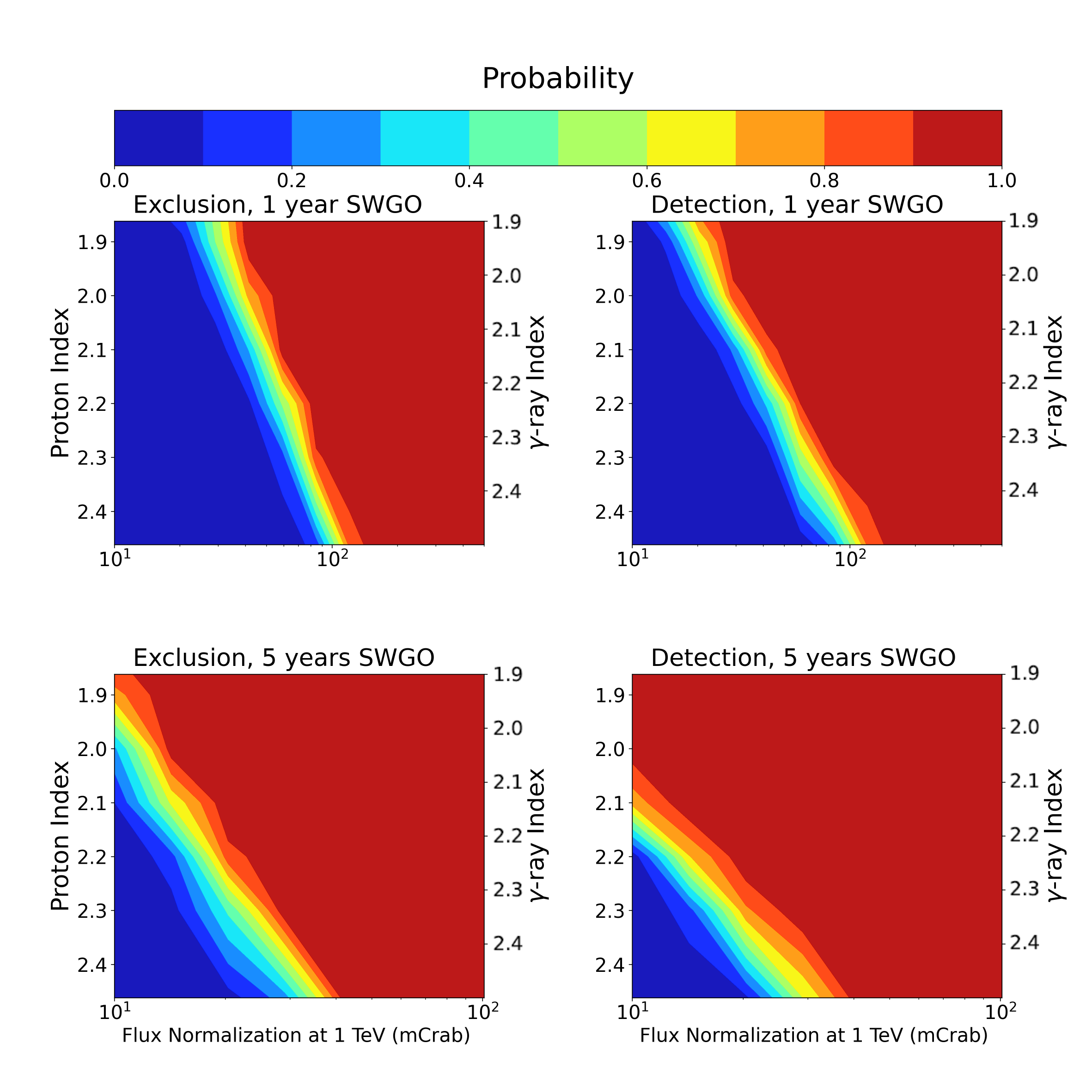}
\caption{Estimated probability for exclusion and detection of a PeVatron association at a respective statistical significance of $5\sigma$ with SWGO data assuming the straw-man \citep{swgo_white} SWGO sensitivity taken from https://github.com/harmscho/SGSOSensitivity. The abscissa, left and right ordinate show the assumed parameters for the true observed $\gamma$-ray flux normalization $\Phi_0$ at $1$~TeV originating from pp interactions observed from Earth, the spectral index of the hadronic particle population $\Gamma_\mathrm{P}$ and the corresponding spectral index of ECPL $\gamma$-ray emission. The color bar indicates the probability to exclude (left panels) and detect (right panels) a PeVatron with a statistical significance of $5\sigma$ with the PTS. One year of SWGO observations is shown in the upper panels, while 5~years of SWGO observations are shown in the lower panels. The assumed hadronic cutoff energy for the left panels, which show the exclusion power at a significance of $5\sigma$, is $E_\mathrm{cut,p}$=$400$~TeV. On the other hand, $E_\mathrm{cut,p}$=$3$~PeV is used for the right panels, which show the respective SWGO PeVatron detection power.\\}
\label{fig_ptspower_1TeV}
\end{figure*}

The PeVatron detection and rejection maps provided in Sec.~\ref{sec_par_scan} uses the flux normalization of sources at 10~TeV, which is much more suited for WCD experiments that have enhanced high energy flux sensitivity. In order to connect these maps to current and future VHE experiments which in general have their maximized flux sensitivity at 1~TeV, the maps reconstructed using $\Phi_{0}$ at 1~TeV are also provided in this appendix in Fig.~\ref{fig_ptspower_1TeV}.

\bigskip

{
{\bf \large List of Acronyms}\\
\\
}
{
\bf{CL: }{ Confidence Level}\\
\bf{CR: }{ Cosmic Ray}\\
\bf{CTA: }{ Cherenkov Telescope Array}\\
\bf{EAS: }{ Extensive Air Shower}\\
\bf{ECPL: }{ Power-Law with Exponential Cutoff}\\
\bf{FoV: }{ Field of View}\\
\bf{GC: }{ Galactic Center}\\
\bf{H.E.S.S.: }{ High Energy Stereoscopic System}\\
\bf{HAWC: }{ High Altitude Water Cherenkov Observatory}\\
\bf{HE: }{ High Energy}\\
\bf{IACTs: }{ Imaging Atmospheric Cherenkov Telescopes}\\
\bf{IRF: }{ Instrument Response Function}\\
\bf{LHAASO: }{ Large High Altitude Air Shower Observatory}\\
\bf{LL: }{ Lower Limit}\\
\bf{MAGIC: }{Major Atmospheric Gamma-Ray Imaging Cherenkov}\\
\bf{MC: }{ Monte-Carlo}\\
\bf{PL: }{ Power-Law}\\
\bf{PSF: }{ Point Spread Function}\\
\bf{PTS: }{ PeVatron Test Statistics}\\
\bf{SN: }{ Supernova}\\
\bf{SNR: }{ Supernova Remnant}\\
\bf{SWGO: }{ Southern Wide-field Gamma-ray Observatory}\\
\bf{TS: }{ Test Statistics}\\
\bf{UHE: }{ Ultra High Energy}\\
\bf{VHE: }{ Very High Energy}\\
\bf{WCD: }{ Water Cherenkov Detector}\\
\bf{YMC: }{ Young Massive stellar Cluster}\\
}

\newpage

\bibliographystyle{elsarticle-harv} 
\bibliography{SWGO_pevatrons}

\end{document}